\documentclass[reprint,amsmath,amssymb,aps,superscriptaddress,longbibliography]{revtex4-1}
\usepackage{dcolumn}% Align table columns on decimal point
\usepackage{bm}% bold math
\usepackage{braket}
\usepackage[dvipdfmx]{graphicx}
\usepackage[dvipdfmx]{color}
\usepackage{pstricks,pst-grad,color}
\usepackage{appendix}
\usepackage{here}
\usepackage{url}
\usepackage{hyperref}
\usepackage{autobreak}
\usepackage{svg}
\usepackage[utf8]{inputenc}
\DeclareUnicodeCharacter{0306}{}

\usepackage{comment}
\usepackage{tikz}
\usetikzlibrary{shapes,calc}
\usetikzlibrary{shapes.geometric,arrows.meta,decorations.markings}

\begin{document}

\title{Unconventional gap dependence of high harmonic generation in the extremely strong light-matter coupling regime}
\author{Akira Kofuji}
\email{kofuji.akira.46c@st.kyoto-u.ac.jp}
 \affiliation{Department of Physics, Kyoto University, Kyoto 606-8502, Japan}
\author{Robert Peters}%
 \email{peters@scphys.kyoto-u.ac.jp}
 \affiliation{Department of Physics, Kyoto University, Kyoto 606-8502, Japan}
\date{\today}% It is always \today, today,

\begin{abstract}
High harmonic generation(HHG) is one of the most commonly studied nonlinear optical phenomena, originating in the ultrafast dynamics of electrons in atomic gasses and semiconductors. It has attracted much attention because of its non-perturbative nature and potential for future attosecond laser pulse sources.
On the theory side, a semi-classical picture based on tunneling ionization of electrons is successfully used in explaining key characteristics of the HHG. 
This model assumes that electric fields non-perturbatively excite electrons beyond the ionization potential or band gap. Thus, intuitively, a larger gap should lead to an exponentially smaller HHG emission.
Despite this intuition, the HHG in the Mott insulator $\mathrm{Ca_{2} Ru O_{4}}$ has shown an unconventional exponential increase with respect to the gap width.
This experiment implies effects beyond the semi-classical theory. However, most theoretical works have focused on the dependence of the HHG on external control parameters, and the gap dependence of the HHG is poorly understood even in non-interacting systems. Thus, it is essential to clarify the gap dependence of the HHG in a fully quantum mechanical approach.
Here, we analyze numerically exactly the gap dependence of the HHG in two-level systems. We find an increase in the strength of the HHG when the Rabi frequency is large compared to the gap width. Furthermore, the relaxation and scattering of electrons increase the visibility of this gap dependence.
Finally, we find that the enhancement rate follows a universal scaling law regardless of the driving frequency.
The existence of this gap dependence in two-level systems suggests that this unconventional gap dependence is a universal behavior that can be found not only in Mott insulators but also in atomic gasses and semiconductors.
\end{abstract}

\maketitle

\section{Introduction}
\label{intro}
Decades of progress in strong laser light technologies have made it possible to probe and control the ultrafast dynamics of electrons in materials.
One of the most commonly studied optical phenomena in the field of strong laser light is the high harmonic generation(HHG) in atomic gases\cite{mcpherson1987studies,ferray1988multiple, krausz2009attosecond} and solids\cite{ghimire2011observation, ghimire2019high}, where photons with multiples of the driving photon energy are emitted, and the spectrum consists of a characteristic plateau and cutoff energy.
HHG has attracted much attention not only because it is a non-perturbative phenomenon with great potential for future attosecond light sources but also because it can be a new all-optical probe for states of matter, i.e., high harmonic spectroscopy\cite{itatani2004tomographic,wagner2006monitoring,baker2006probing,patchkovskii2009nuclear,worner2011conical,le2012theory,vampa2015all,luu2015extreme,hohenleutner2015real,lein2005attosecond,you2017anisotropic,kaneshima2018polarization,luu2018measurement,lakhotia2020laser,uchida2021visualization,bionta2021tracking,heide2022probing,bae2022revealing,rana2022high}.
For example, HHG is now utilized for probing molecular orbitals\cite{itatani2004tomographic}, band structures\cite{vampa2015all}, Berry curvature of materials\cite{luu2018measurement,bae2022revealing}, topological phase transitions\cite{heide2022probing}. Recently, various types of HHG, unique to strongly correlated electron systems, have also been intensively studied\cite{liu2017high,silva2018high,murakami2018high,takayoshi2019high,imai2020high,lysne2020signatures,murakami2021high,zhu2021ultrafast,orthodoxou2021high,shao2022high,alcala2022high,hansen2022correlation,uchida2022high,alshafey2022ultrafast,pizzi2023light,shimomura2023ultrafast}, opening possibilities to detect ultrafast light-induced phase transitions\cite{silva2018high}, quantum phase transitions\cite{shao2022high}, and transitions from the strange metal phase to the pseudogap phase\cite{alcala2022high}.

Almost all of these previous studies can be understood based on the well-known semiclassical 3-step model\cite{Corkum1993,lewenstein1994theory}, which consists of (1) tunnel ionization of electrons trapped in some potential, (2) forced oscillatory motion, and (3) recombination of electrons in the potential.
This simple yet powerful model has had great success in explaining the most important characteristics of the HHG, i.e., the plateau and the cutoff energy, and has been extended to solids\cite{vampa2015}.
This model leads us to the naive expectation that the tunneling amplitude becomes exponentially small when the excitation gap or the depth of the potential becomes large. Thus, the HHG emission should also become exponentially small.

Recently, an unconventional HHG which contradicts this intuition has been experimentally observed in the Mott insulator, $\mathrm{Ca_{2} Ru O_{4}}$\cite{uchida2022high}. Utilizing the temperature dependence of the Mott gap of this material, the dependence of the HHG on the excitation gap has been investigated, and it has been shown that the HHG grows exponentially as the gap width increases.
More surprisingly, the gap dependence obeys an empirical and universal scaling law regardless of the driving frequency.
While such a strange gap dependence has been observed in  experiments, most theoretical studies have focused on the dependence of the HHG on external control parameters, such as the ellipticity, strength, and carrier-envelope phase of the incident laser pulse\cite{yoshikawa2017high,Ishii2014}.
Recently, Murakami {\it et al.}\cite{Murakami2022anomalous} have studied the unconventional temperature (gap) dependence of the HHG by analyzing the laser-driven Hubbard model with non-equilibrium dynamical mean field theory, suggesting a relation between the unconventional gap dependence and doublon scattering. 
Murakami {\it et al.}\cite{murakami2022doping} have also studied the gap dependence of the HHG in gapped graphene, where they have found that the modification of the intra-band dipole via inter-band transitions leads to a non-monotonic gap dependence of the HHG.
While these studies show that the HHG can have an unconventional gap dependence in correlated systems and gapped graphene, the gap dependence of the HHG in non-interacting systems is actually not well understood. It remains unclear to what extent the unconventional gap dependence is due to correlations and whether it can also be observed in uncorrelated systems.
Furthermore, understanding the gap dependence of the HHG can open a new path to realize a stronger HHG emission by tuning the excitation gap of materials and new high harmonic spectroscopic methods through the gap dependence of the HHG.

We here demonstrate that an unconventional gap dependence of the HHG can be observed even in uncorrelated models. 
For this purpose, we study the HHG in a two-level system.
Two-level systems are very simplified models that do not include intraband current and multiband effects, which occur in realistic solid-state systems.
However, two-level systems still share various aspects of the HHG in atomic gases and solids\cite{kaplan1994superdressed,krainov1994plateau,de2002high}.
The advantage of these simplified models is that we can reduce the numerical cost significantly. Thus, we can investigate the HHG in a fully quantum mechanical approach and numerically exactly in a wide range of parameters, such as the Rabi frequency, relaxation time, and the excitation gap.
Based on this two-level system, we first analyze the gap dependence of the HHG varying the Rabi frequency. 
We observe an increase in the HHG with increasing gap width when the Rabi frequency is large compared to the gap. In contrast, the HHG decreases exponentially when the gap is large.
To investigate the origin of this gap dependence, we consider the effects of relaxation processes on the HHG and calculate the time-resolved spectrum of the HHG.
We see that relaxation is not necessary for the appearance of an unconventional gap dependence. 
Finally, we study the enhancement ratio at each emission energy and see that it obeys a universal behavior regardless of the incidental frequency, which is also observed in the experiment\cite{uchida2022high}. 
From the above analysis, we expect that this unconventional gap dependence of the HHG, as observed in the experiment, is a universal characteristic not only found in Mott insulators but also in atomic gases and semiconductors. 

The remainder of this paper is structured as follows:
Section~\ref{model_and_methods} introduces the two-level system driven by an electric laser pulse.
In Sec.~\ref{results}, we analyze the gap dependence of the HHG and the effects of relaxation processes. 
Finally, we calculate the enhancement ratio of the emission energy and show its universal behavior.
Finally, we conclude the paper in Sec.~\ref{conclusion}.
Furthermore, we show an analysis of the HHG in semiconductors based on semiconductor Bloch equations in  appendix~\ref{appenxix_semiconductor}. 

\section{Model and Methods}
\label{model_and_methods}
First, we explain the Hamiltonian of the two-level system. In the paper, we use the following units: $c=e=\hbar=d_{12}=1$, which correspond to the speed of light, the elementary charge, Planck's constant, and the transition dipole moment of the system.
Then, the Hamiltonian can be written as
\begin{equation}
\label{hamiltonian}
\hat{H}(t) = \hat{H}_{0} + \hat{H}_\text{ext} = 
\begin{bmatrix}
-\Delta/2 & \Omega(t) \\
\Omega(t) & \Delta /2 \\
\end{bmatrix}.
\end{equation}
$\hat{H}_{0}$ is the original Hamiltonian of the two-level system, which corresponds to the diagonal part of $\hat{H}$. $\hat{H}_\text{ext} (t)$ is the contribution from the external electric field, which corresponds to the off-diagonal part of $\hat{H}$. $\Delta$ represents the gap of the system, and $\Omega(t)$ depends on the strength and the shape of the electric field.

To take relaxation processes into account, we use the von Neuman equation, which is more suitable for calculating the time evolution of this system.
Calculating the commutator of the Hamiltonian, Eq.~(\ref{hamiltonian}), and the density matrix of the system, $\hat{\rho}$, and adding relaxation terms, we arrive at the following equation of motion\cite{kaplan1994superdressed}:
\begin{eqnarray}
\label{equation_of_motion}
\dot{x} &=& 2 i \Omega(t) (y-y^{*}) - \gamma_{l}(x-1)\\
\dot{y} &=& - \left(i \Delta + \gamma_{t} \right) y +i \Omega(t) x .
\end{eqnarray}
$x=\rho_{11}-\rho_{22}$ is the population difference of the levels in the system, and $y=\rho_{21}$ measures the coherence between both levels.
$\gamma_{l}$ is the longitudinal relaxation rate, which induces the relaxation from the upper level to the lower level, and $\gamma_{t}$ is the transverse relaxation rate, which induces decoherence between both levels.
The origin of these relaxation terms can be traced back to spontaneous emission, inter-atomic interactions, electron-electron interactions, and electron-phonon interactions. In this paper, it is not our primary purpose to study how these relaxation terms appear from microscopic interactions. Thus, we include these relaxation processes just phenomenologically.
(For details, see Ref.~\cite{boyd2020nonlinear}.)

In two-level systems, the origin of the HHG is a fast oscillation of the polarization of the system. To obtain the spectrum of the HHG, first, we numerically calculate the time dependence of the density matrix $\rho$ of the system and obtain the time-dependent expectation value of the position operator, corresponding to the polarization given as
\begin{equation}
p(t) = \mathrm{Tr} [\hat{\rho}(t) \hat{x}] = y+y^{*}.
\end{equation}
Then, by calculating the Fourier transform of the polarization, we obtain the spectrum of the HHG. We note that the transition dipole moment is assumed to be constant here.
The external field is modulated by a Gaussian amplitude and is written as
\begin{equation}
\Omega(t) = \Omega_{0} \cos (\omega_{0} t) e^{-t^{2}/\tau^{2}}.
\end{equation}
$\Omega_{0}$ is the Rabi frequency, which is the product of the strength of the electric field and the transition dipole moment. $\omega_0$ is the frequency of the incidental light.

\begin{figure}[t]
\begin{center}
\includegraphics[width=\linewidth]{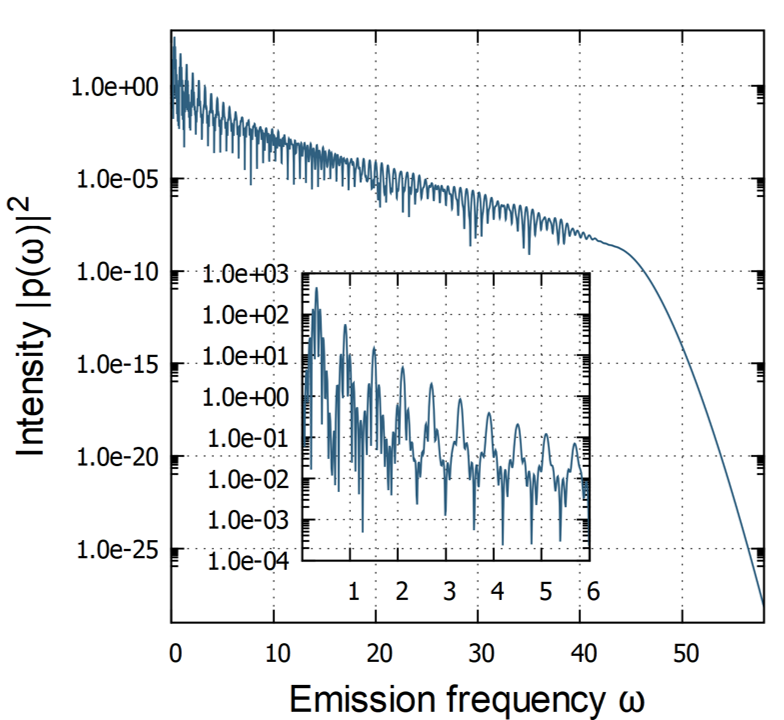}
\end{center}
\caption{A typical example of an HHG spectrum. The inset shows the enlarged view of the spectrum for low frequencies. The parameters are $\Delta=2.1$, $\Omega_{0}=22.7$, $\gamma_{l}=2.0$, $\gamma_{t}=1.0$, $\tau=8.5\pi$, $\omega_{0}=0.3$. The vertical axis is log-scale.}
\label{spectrum_example}
\end{figure}

The HHG in a two-level system has common characteristics with the HHG in atomic potentials and solids. We here briefly review the mechanism of the HHG in two-level systems following Ref.~\cite{de2002high}.
For the sake of simplicity, we ignore relaxation processes and analyze the Hamiltonian in Eq.~(\ref{hamiltonian}).
We note that the "adiabatic" basis is more convenient for understanding the origin of the HHG.
The adiabatic basis diagonalizes the time-dependent Hamiltonian $\hat{H}(t)$ at each instant of time.
This can be done by the following unitary operator,
\begin{equation}
\hat{U}(t) = 
\begin{bmatrix}
\cos (\chi(t)) & \sin (\chi(t)) \\
-\sin (\chi(t)) & \cos (\chi(t)) \\
\end{bmatrix},
\end{equation}
where $\chi(t) = \frac{1}{2} \tan^{-1} (\frac{2\Omega(t)}{\Delta})$.
In this basis, the time evolution of the system is described by the following Hamiltonian,
\begin{equation}
\hat{H}'(t) = {\hat{U}}^{\dagger}(t) \hat{H}(t) \hat{U}(t) + i \frac{\partial \hat{U} (t)}{\partial t} = 
\begin{bmatrix}
    \varepsilon_{-}(t) & i \dot{\chi} \\
    -i \dot{\chi} & \varepsilon_{+}(t) \\
\end{bmatrix},
\end{equation}
where $\varepsilon_{\pm} = \pm \frac{1}{2} \sqrt{\Delta^{2}+4\Omega^{2}(t)}$ are the "eigenenergies" at each instant, and $i\dot{\chi}$ induces a non-adiabatic transition between the "energy" levels, which is given as $\dot{\chi} = \frac{\dot{\Omega}/\Delta}{1+(2\Omega(t)/ \Delta)^{2}}$.
Transitions between both states can easily occur near the nodes of the pulse, where $\Omega(t)=0$.
On the other hand, the states evolve adiabatically 
near the antinodes of the pulse, where $\Omega(t)$ is large.
Thus, in intervals of time with small electric fields, the electrons in the lower "energy" level are excited to the upper level, then perform an adiabatic time evolution generating ultrafast oscillations of the polarization, and finally return to the lower level. 
This process is analogous to the three-step model in atomic gases and solids. Furthermore, in semiconductors, the description of the system is identical to that of the two-level system under special conditions, such as low doping and identical effective masses in the valence and conduction bands\cite{birnir1993nonperturbative,nikonov1997collective,de2002high}. Therefore, the HHG in two-level systems can capture the general characteristics of the HHG in various systems.

A typical example of an HHG spectrum, calculated in the two-level system, is shown in Fig.~\ref{spectrum_example}.
The spectrum shows the characteristic of the HHG, i.e., the plateau, at which the intensity of the spectrum stays constant or only decreases slightly with increasing harmonic order, and the cutoff at high frequencies, above which the intensity decreases quickly.
We note that throughout the paper, the incidental frequency is much smaller than the gap energy. Thus, the system is off-resonant.
We consider the case in which the tunneling ionization is dominant compared to multi-photon excitations.

\section{Results}
\label{results}
\begin{figure}[t]
\begin{center}
\includegraphics[width=\linewidth]{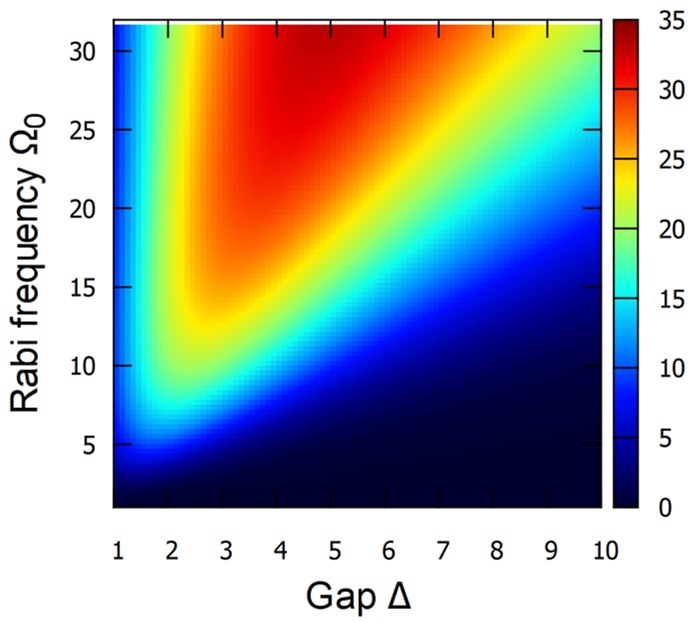}
\end{center}
\caption{Gap and Rabi frequency dependence of the 5th harmonics ($\omega=5\omega_{0}$).}
\label{gap_field_dependence_color}
\end{figure}
\begin{figure}[t]
\begin{center}
\includegraphics[width=0.9\linewidth]{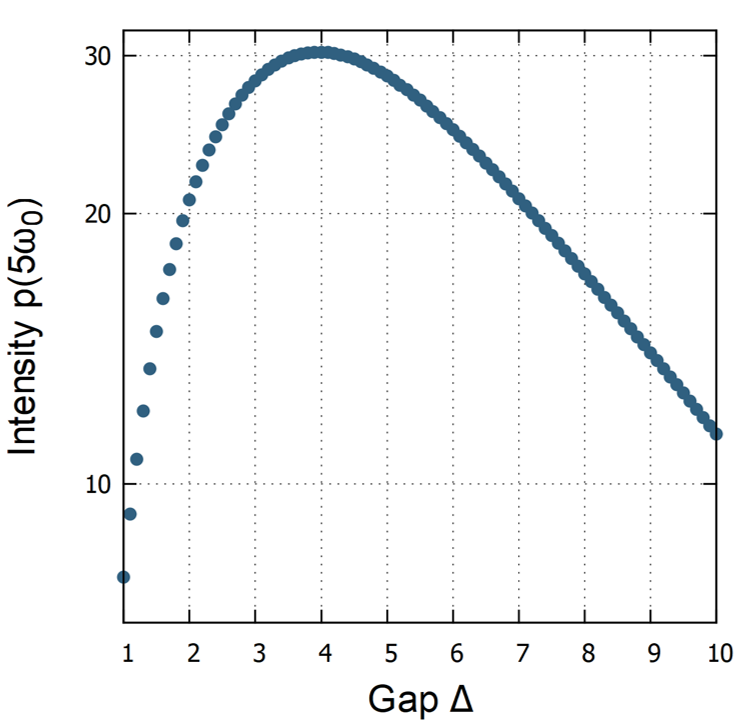}
\end{center}
\caption{ Gap dependence of the 5th harmonics for $\Omega_{0}=22.7$. The vertical axis is log-scale.}
\label{gap_field}
\end{figure}
\subsection{Unconventional gap dependence}
\begin{figure*}[t]
\begin{center}
\includegraphics[width=0.32\linewidth]{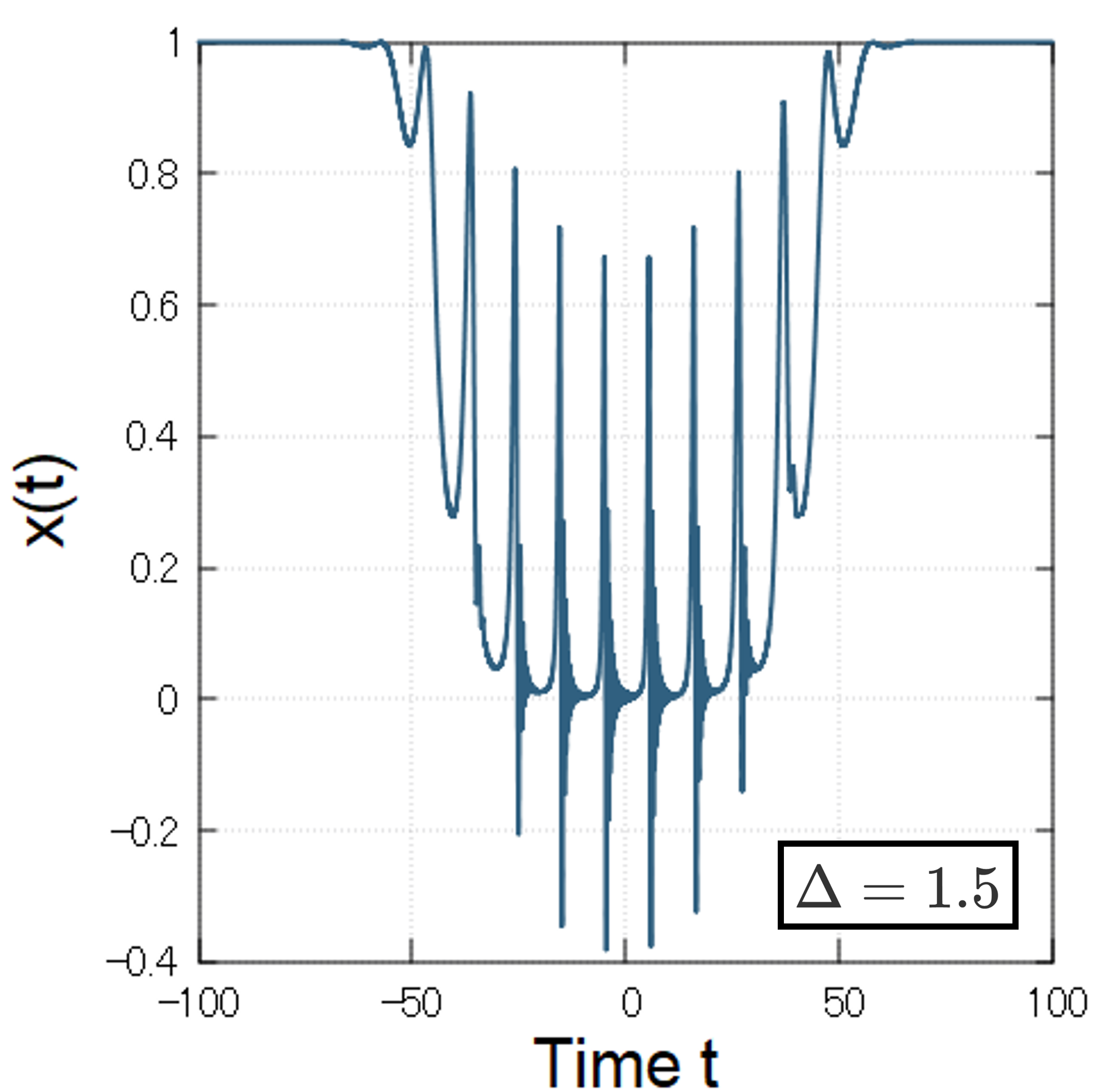}
\includegraphics[width=0.32\linewidth]{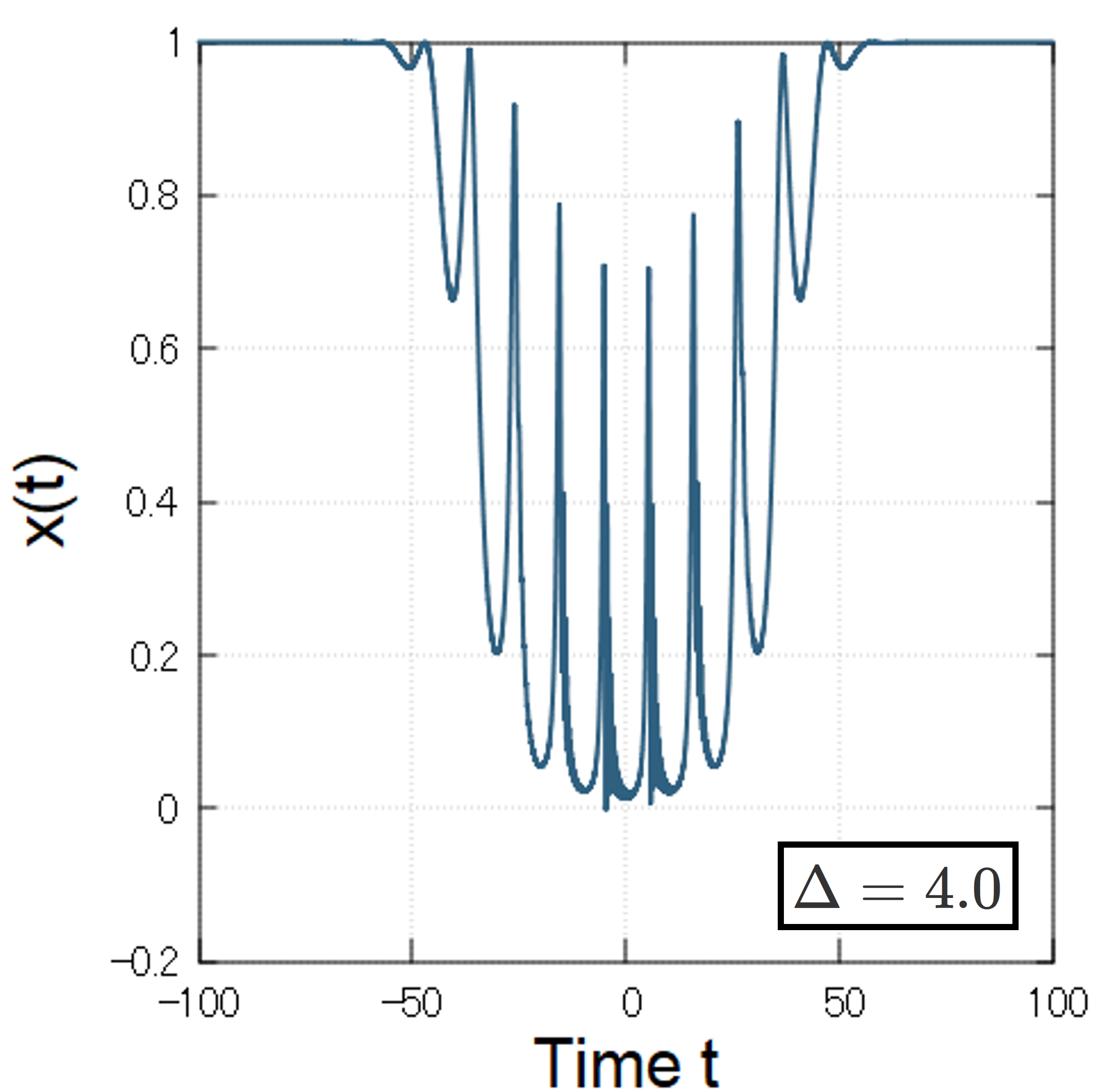}
\includegraphics[width=0.32\linewidth]{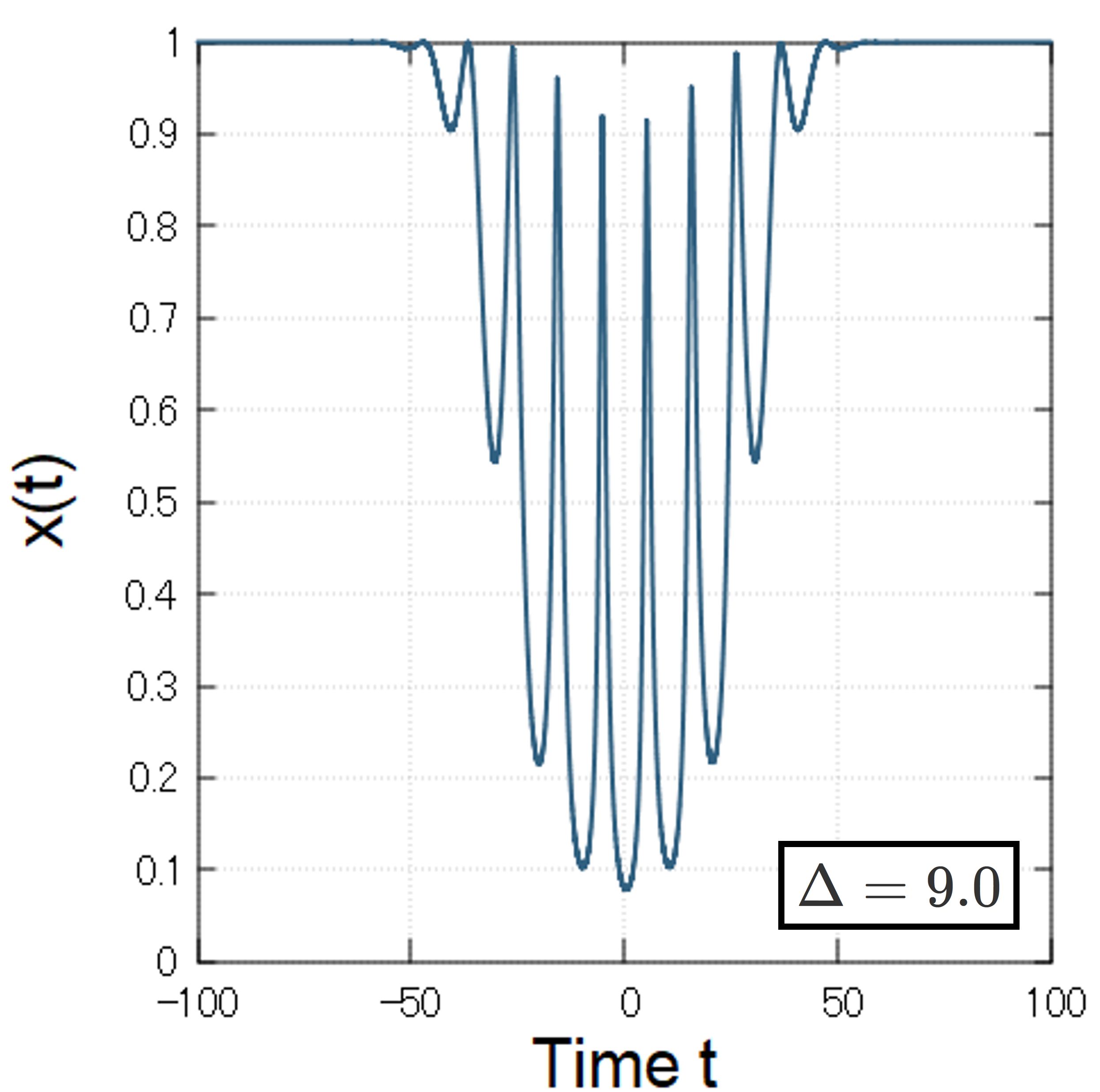}
\end{center}
\begin{center}
\includegraphics[width=0.32\linewidth]{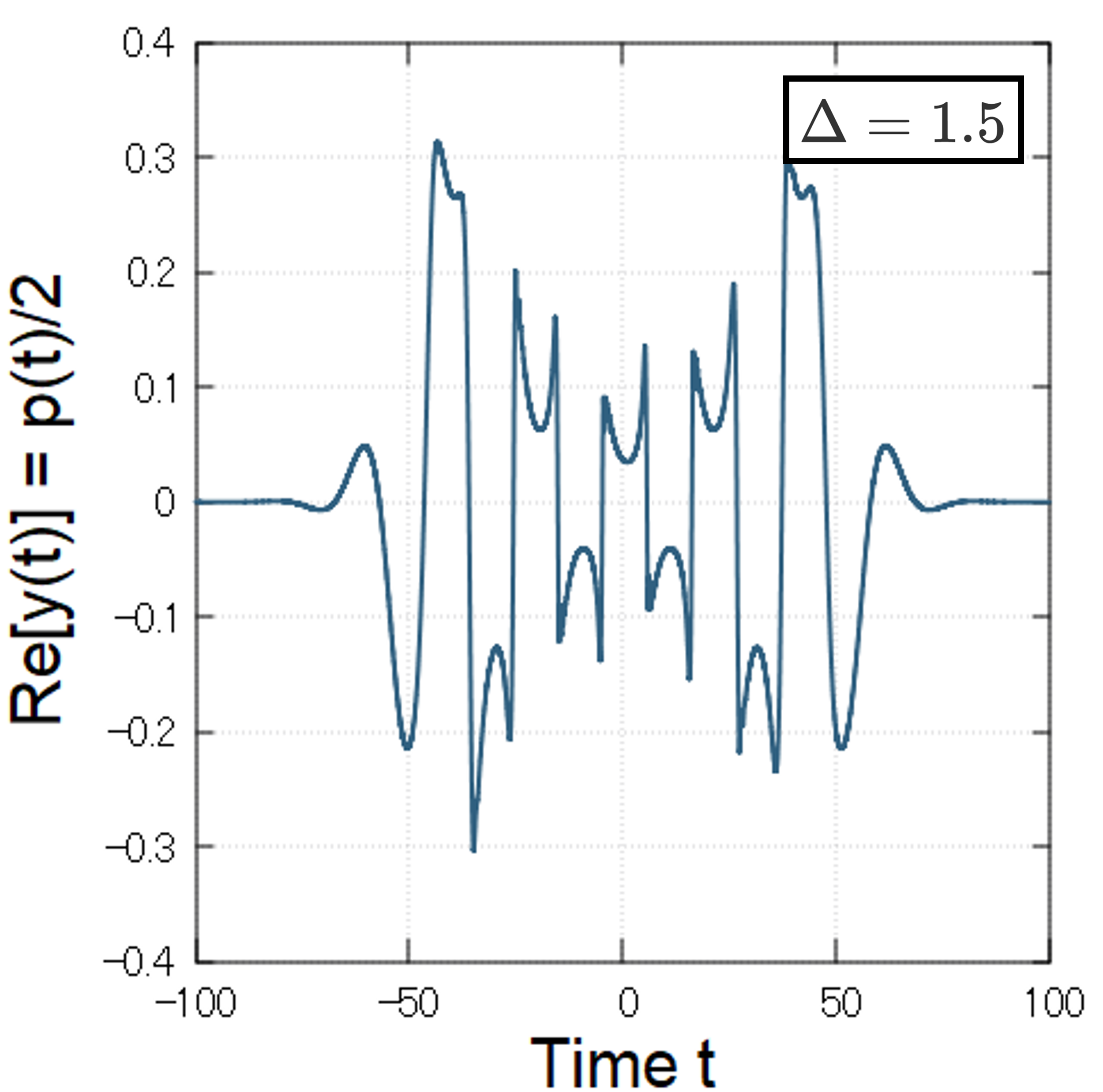}
\includegraphics[width=0.32\linewidth]{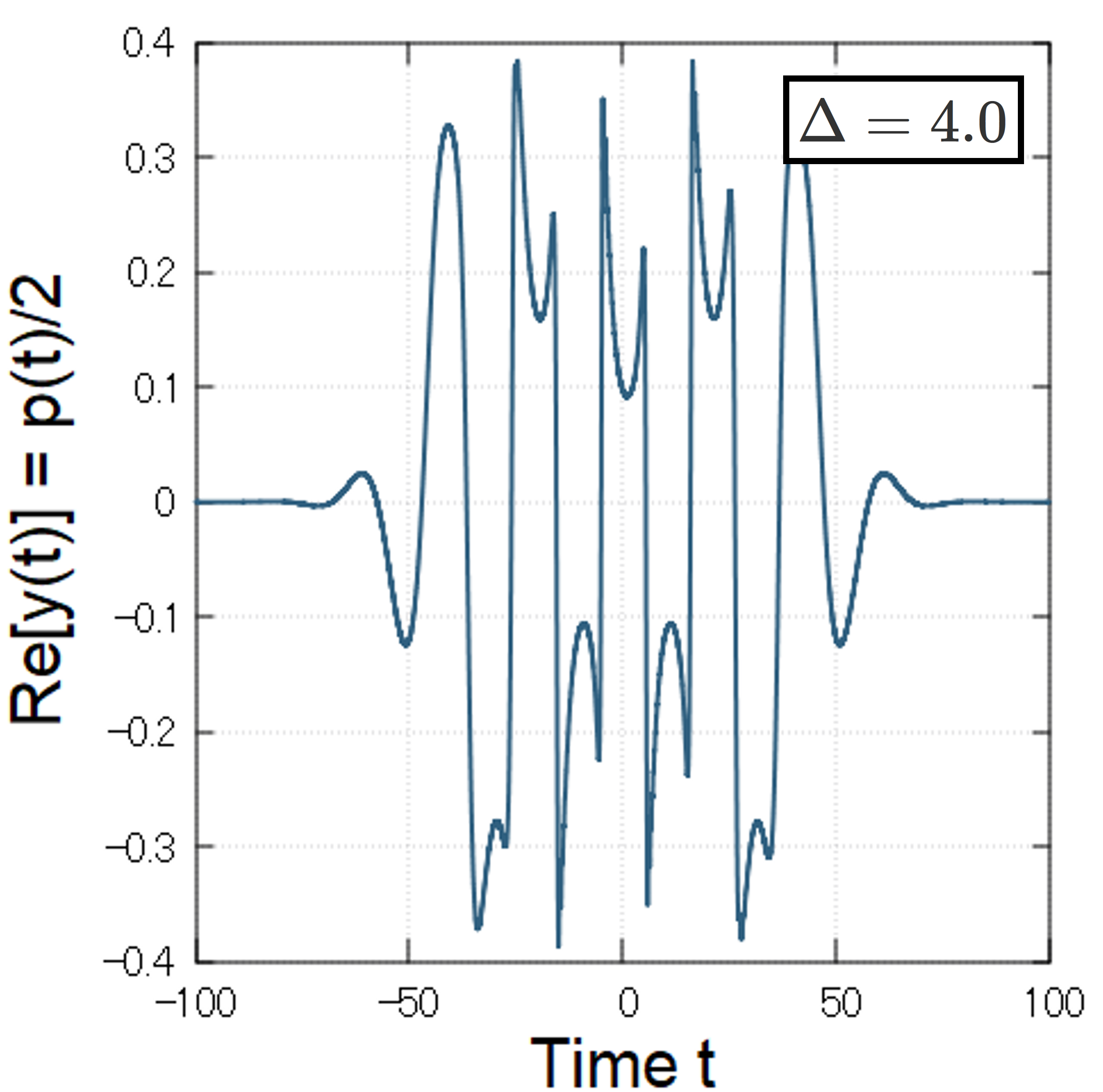}
\includegraphics[width=0.32\linewidth]{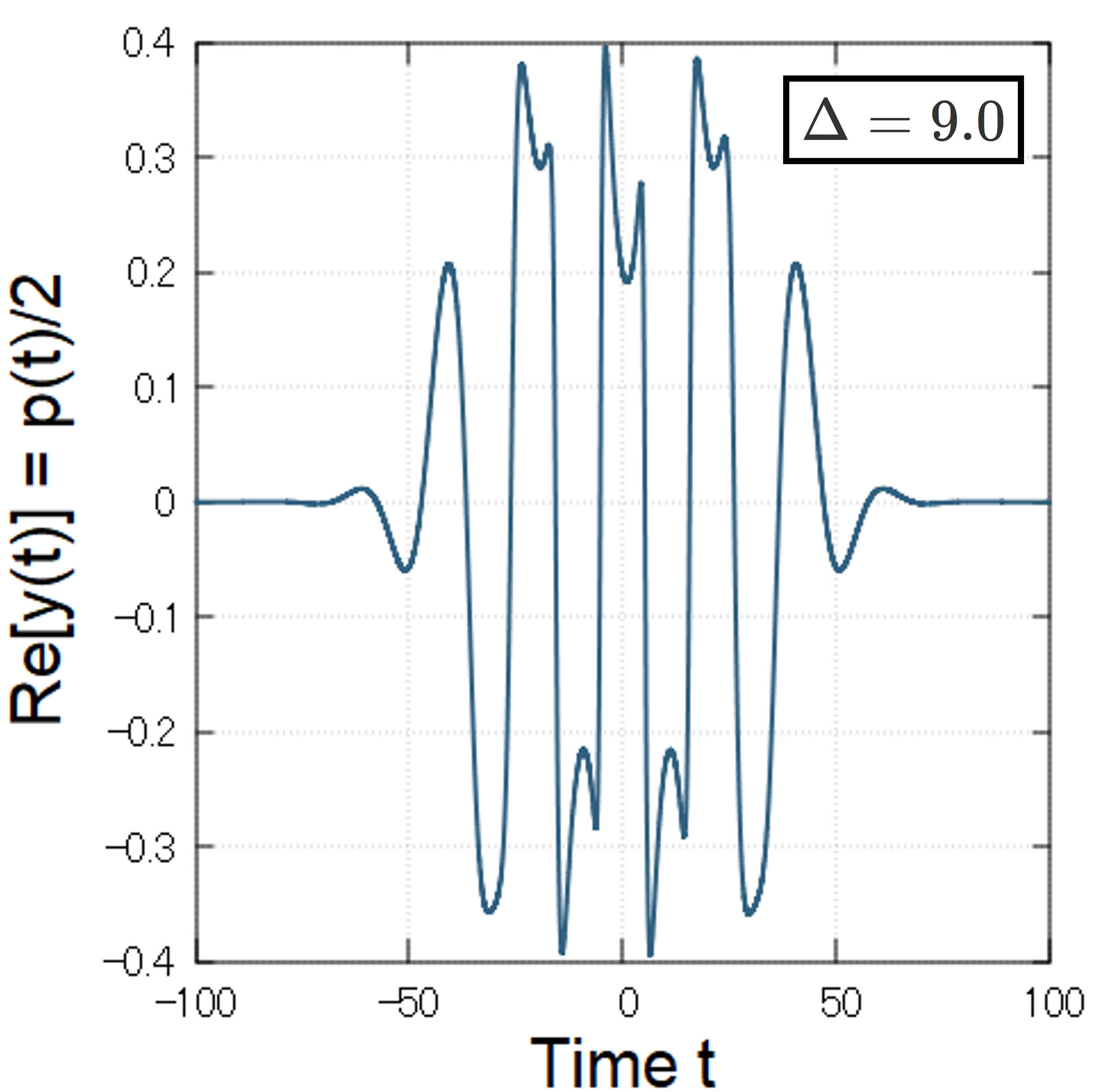}
\end{center}
\caption{Time-resolved $x(t)$ and $\text{Re}(y(t))$ for three different regimes, small gap (left), an intermediate gap where the $5\omega$ HHG is maximal (middle), and large gap (right). The Rabi frequency is $\Omega_{0}=22.7$.}\label{time_resolved_full}
\end{figure*}
First, we analyze the strength of the HHG depending on the gap width, $\Delta$, and the Rabi frequency, $\Omega_0$. 
In this subsection, $\tau$, $\omega_{0}$, $\gamma_{l}$, and $\gamma_{t}$ is fixed as $\tau=8.5\pi$, $\omega_{0} = 0.3$ and $\gamma_{l}=2\gamma_{t}=2.0$, and we vary $\Delta$ and $\Omega_{0}$.
In Fig.~\ref{gap_field_dependence_color}, we show the intensity of the $5\omega_0$-harmonics for various gap widths and Rabi frequencies. 
For sufficiently large values of the Rabi frequency compared to the gap width, i.e., in the upper left region of Fig.~\ref{gap_field_dependence_color}, the intensity of the HHG grows as the gap increases. The intensity takes a maximum at approximately $\Omega_0 / \Delta \sim 6$ and decreases for larger gap widths.
To confirm this behavior, we show the intensity for $\Omega_0=22.7$ over the gap width in Fig.~\ref{gap_field}.
This figure reveals that the intensity %exponentially 
increases until the gap reaches some threshold value and then decreases exponentially. 
This non-monotonic behavior suggests that the naive intuition that larger gaps result in a smaller HHG intensity is not true for large values of the Rabi frequency.
\begin{figure*}[t]
\begin{center}
\includegraphics[width=0.32\linewidth]{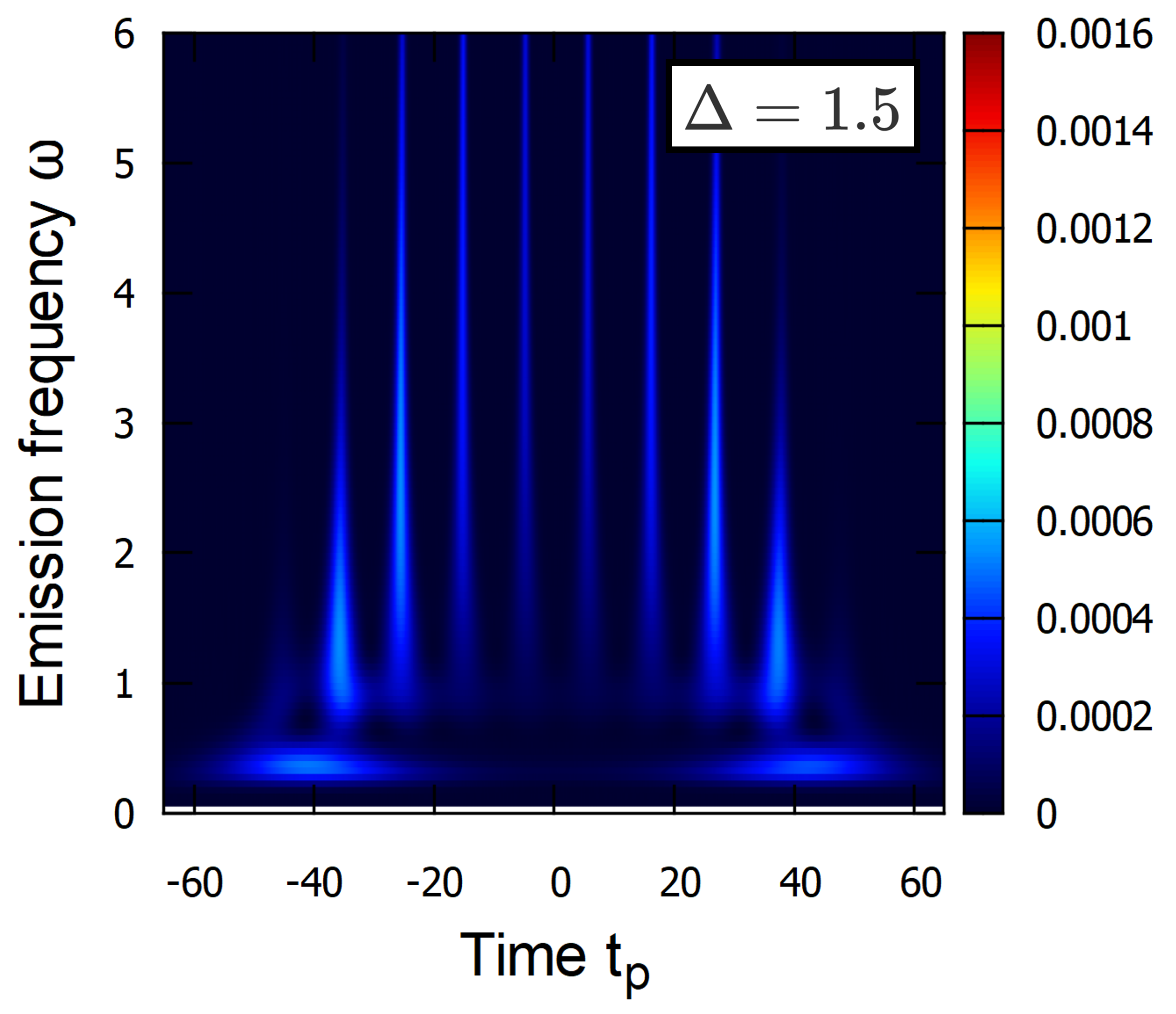}
\includegraphics[width=0.32\linewidth]{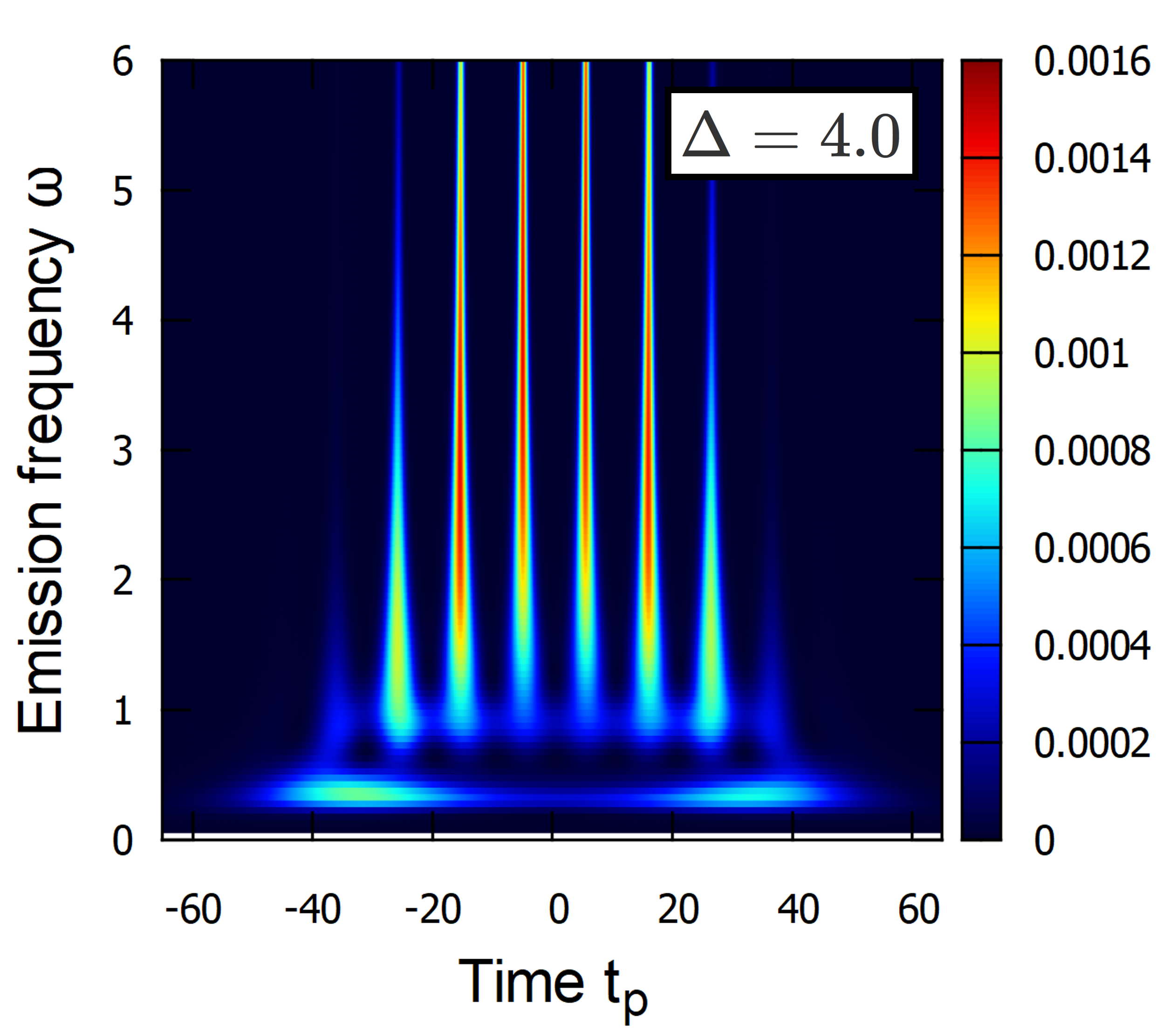}
\includegraphics[width=0.32\linewidth]{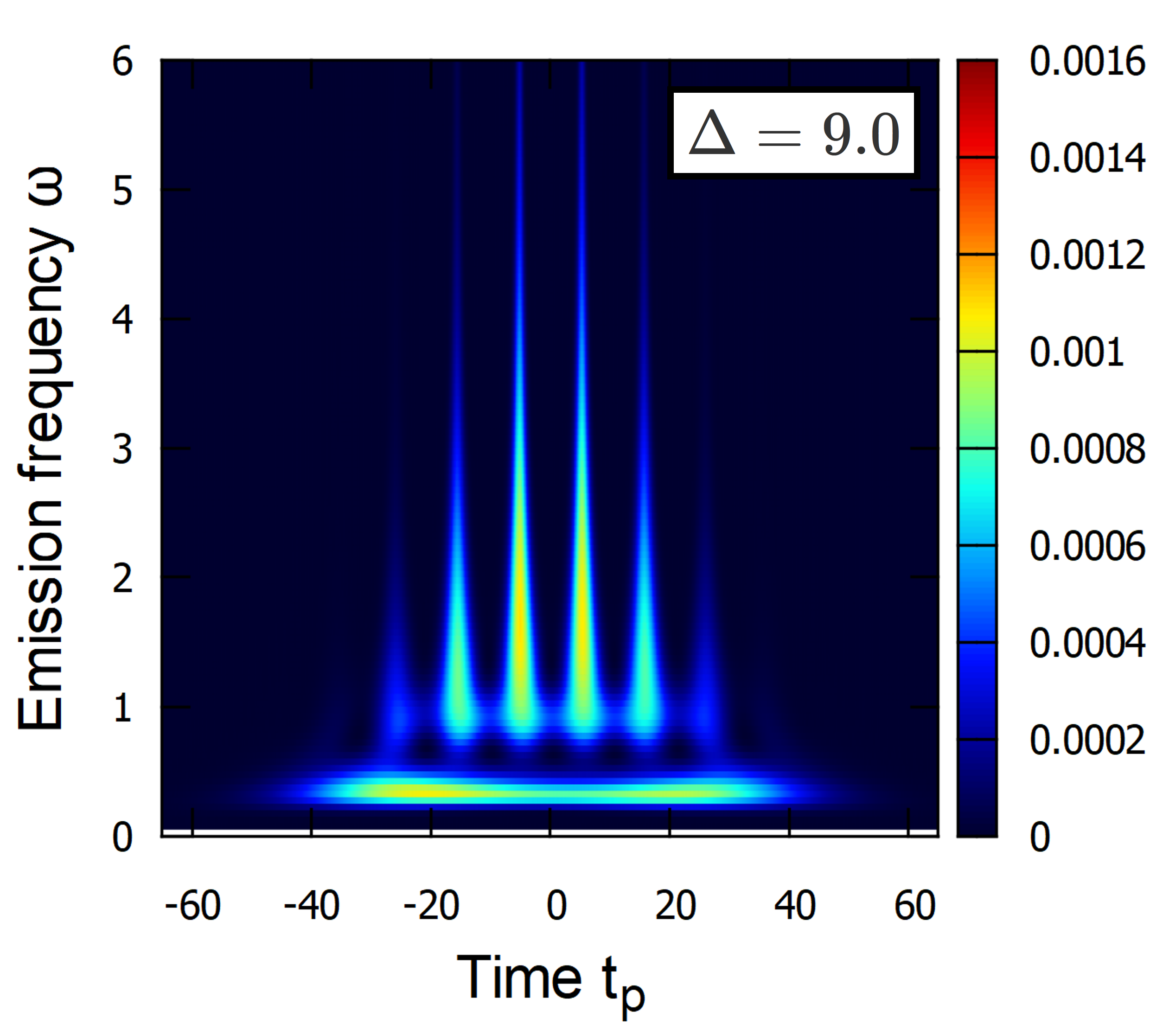}
\end{center}
\caption{
Time-resolved spectrum of HHG for three different gaps(Left: $\Delta=1.5$, Centre: $\Delta=4.0$, Right: $\Delta=9.0$). The Rabi frequency is $\Omega_{0}=22.7$.}
\label{time_resolved_Fourier}
\end{figure*}

To understand the enhancement of the HHG for small gap widths better, we next analyze the time dependence of the HHG.
We first show in Fig.~\ref{time_resolved_full} the time evolution of $x(t)$ and $\mathrm{Re}[y(t)]$ for three different parameters using $\Omega_0=22.7$, corresponding to the three different regimes, i.e., (a) the strength of the HHG increases with increasing gap width at $\Delta=1.5$, (b) the  regime where the strength of the HHG is near the maximum at $\Delta=4$, and (c) the regime where the strength of the HHG decreases with increasing gap width at $\Delta=9$. As described above, $x(t)$ is the occupation difference between both levels and the real part of $y(t)$ corresponds to the polarization.

In all regimes, $x(t)$ decreases at small times; electrons are excited to the other level.
However, we see that $x(t)$ decreases faster for small gaps.
This also results in faster growth of the polarization for small gaps at small times.
This corresponds to the fact that electrons are easily excited when the gap is small.
We can determine the time at which a high harmonic is generated by multiplying the polarization with a Gaussian windowing function before Fourier transform. 
This results in an HHG spectrum, which depends on time through the windowing function as
\begin{eqnarray}
    p(\omega,t_{p}) = \int_{-\infty}^{\infty} dt p(t) W(t,t_{p}) e^{i\omega t} \\
    W(t,t_{p}) = \frac{1}{\sqrt{2 \pi \sigma^{2}}} e^{-\frac{(t-t_{p})^{2}}{2 \sigma^{2}}}
\end{eqnarray}
 The width of the windowing function used here is $\sigma = 3.5$.
 In Fig.~\ref{time_resolved_Fourier}, we show the time-resolved HHG spectrum for 
 the same parameters as in Fig.~\ref{time_resolved_full}.
This figure shows that a reasonably strong HHG is created already at $t=-40$ for $\Delta=1.5$, while the HHG spectrum for $\Delta=4$ and $\Delta=9$ is very small at this time.
This HHG can be seen in the polarization as an additional dip (inside the maximum) for $\Delta=1.5$ at $t=-40$, which is absent for $\Delta=4$ and $\Delta=9$.
Thus, large gaps generally slow down the process of exciting electrons and the generation of high harmonics in the polarization at small times. 

On the other hand, the fast decrease of $x(t)$ at small times due to a large Rabi frequency and a small gap width leads to an extended period in which $x(t)$ is small (besides a periodic spike). This results in a saturation of polarization because 
a finite population difference $x(t)$ is essential for the buildup of polarization (see Eq.~(\ref{equation_of_motion})).
Finally, the magnitude of the polarization even decreases towards the center of the pulse, $t=0$, for $\Delta=1.5$, as can be seen in Fig.~\ref{time_resolved_full}. As a consequence, the contribution to the HHG from the center of the pulse becomes weak. Thus, we can say that for $\Delta=1.5$, the system is too strongly excited, which results in a situation where the HHG is only generated at the beginning and the end of the pulse but not over an extended period in the center of the pulse.

For $\Delta=4$ in Fig.~\ref{time_resolved_full}, we see that the magnitude of the polarization quickly reaches a large value and remains large during the whole pulse. Although the generation of the HHG starts at a later time than for $\Delta=1.5$, the HHG is generated over a longer time period and especially around $t=0$. For $\Delta=9$, the excitations of a large number of electrons, the buildup of a large polarization, and the generation of HHG take a longer time. A large HHG is only generated around the center of the pulse at $t=0$.

We conclude here that these three regimes have not only a different HHG dependence on the gap width but can also be distinguished from their time-dependent polarization, $y(t)$.
Decreasing $\Omega_{0} / \Delta$, we see that 
 the dominant contribution gradually moves from the start/end to the center of the pulse
 In the region where the gap dependence shows a conventional exponential decrease, the dominant contribution comes from the center of the pulse.
On the other hand, in the region where the HHG  grows with an increasing gap, the dominant contribution originates from the start/end of the pulse.
This implies that the origin of the unconventional gap dependence is a very strong light-matter coupling that results in the saturation of the polarization before the center of the pulse, and, thus, the contributions from the center of the pulse become weak.

\subsection{Effects of relaxation on the gap dependence}
\label{relaxtion}
\begin{figure*}[t]
\begin{center}
\includegraphics[width=0.32\linewidth]{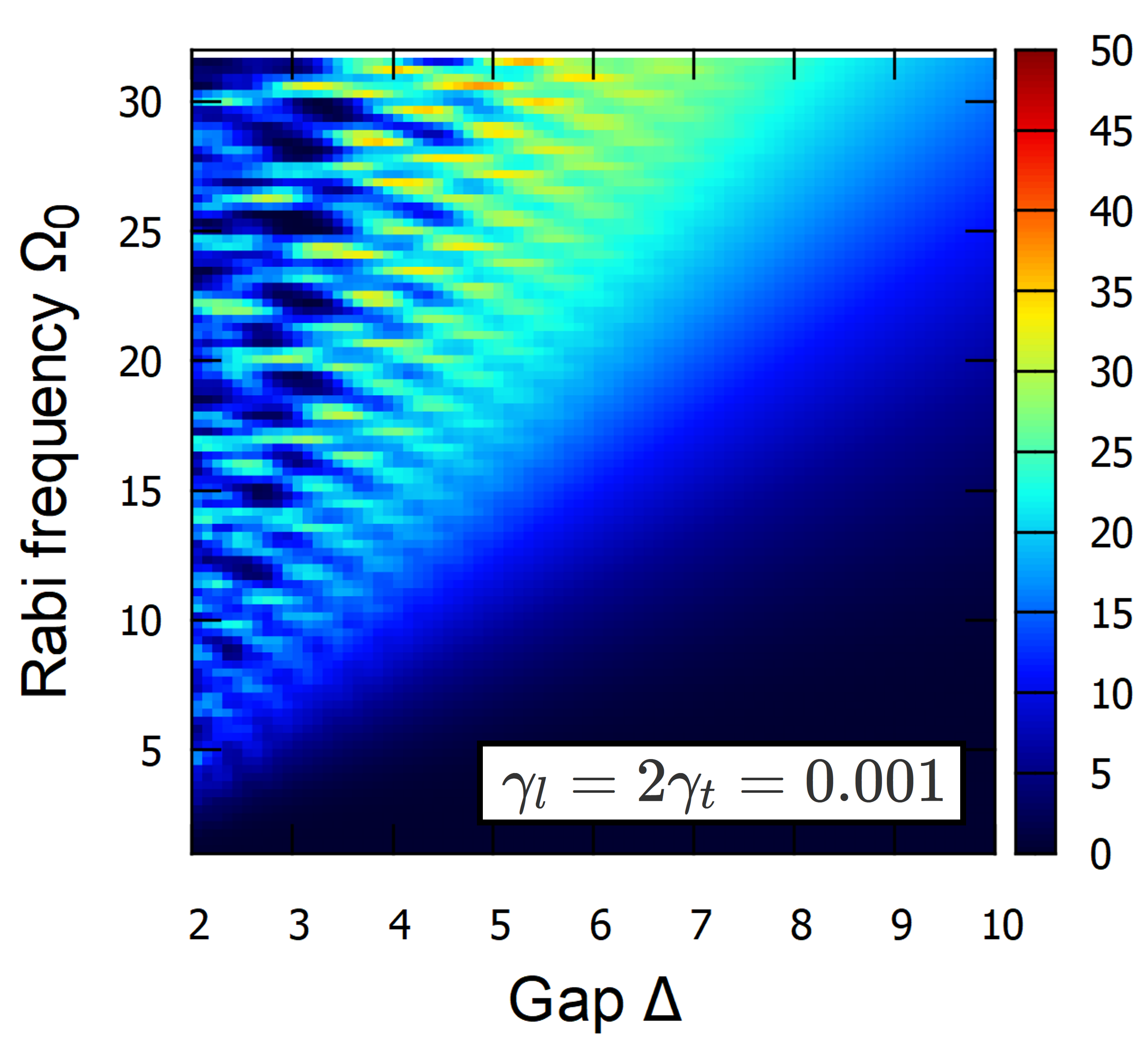}
\includegraphics[width=0.32\linewidth]{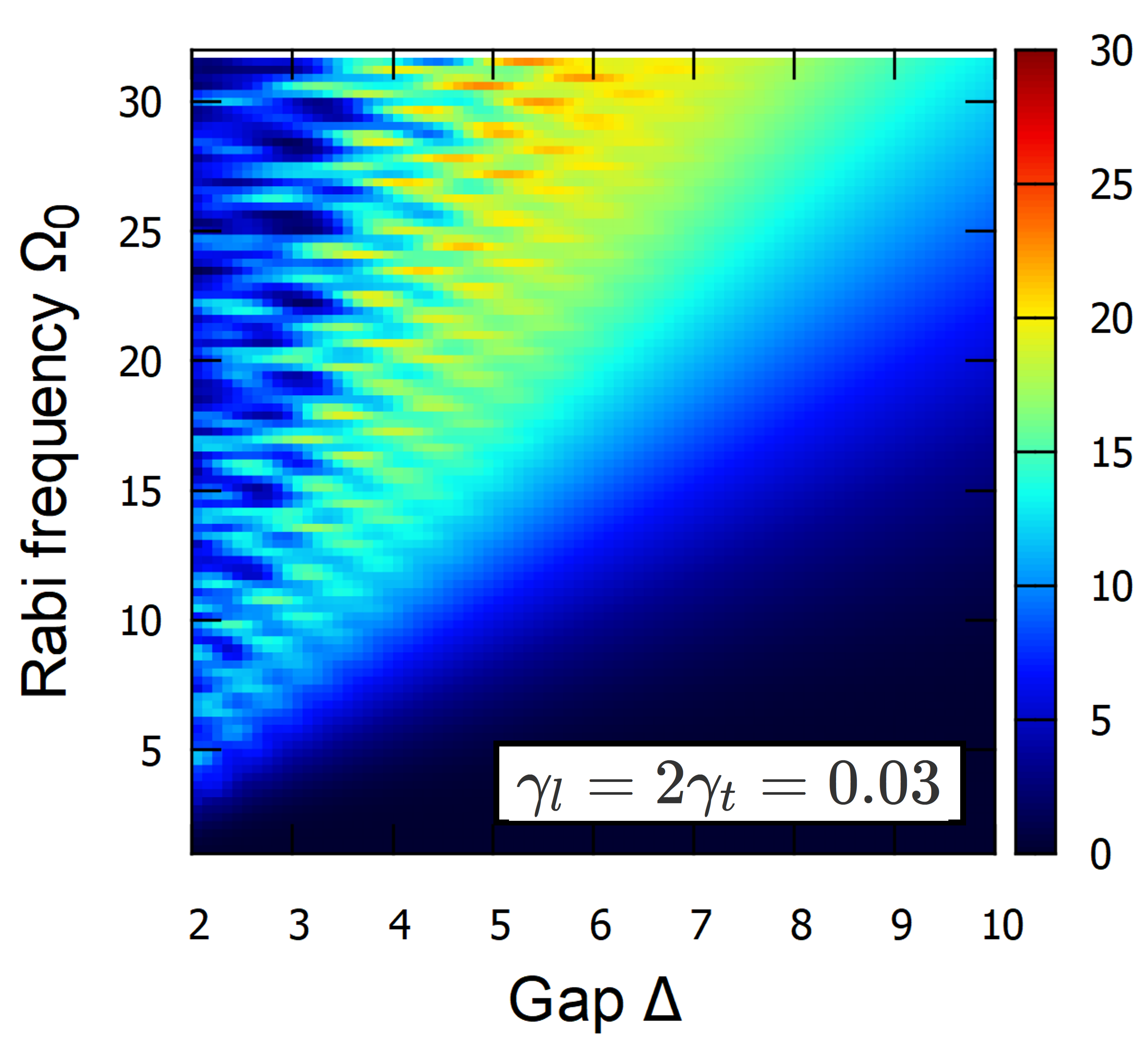}
\includegraphics[width=0.32\linewidth]{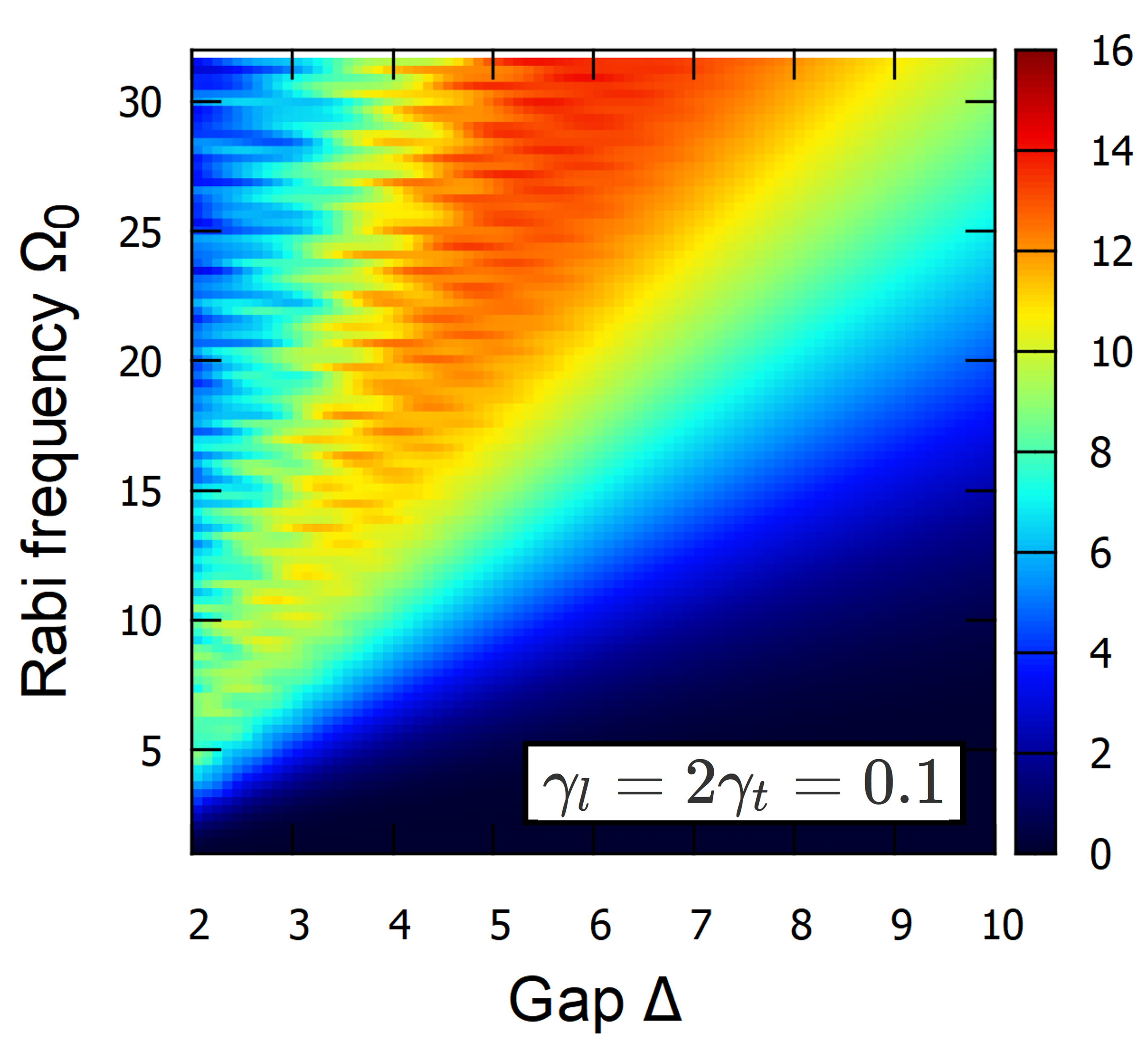}
\end{center}
\caption{Gap and Rabi frequency dependence of the HHG for three different relaxation rates (left: $\gamma_{l}=2\gamma_{t}=0.001$, center: $\gamma_{l}=2\gamma_{t}=0.03$, and right: $\gamma_{l}=2\gamma_{t}=0.1$).}
\label{relax}
\end{figure*}
Next, we study the effect of relaxation on the HHG. 
In this subsection, $\tau$ and $\omega_{0}$ is fixed as $\tau=8.5\pi$ and $\omega_{0}=0.3$, and we vary $\Delta$, $\Omega_{0}$, $\gamma_{l}$, and $\gamma_{t}$.
Relaxation processes are common in various systems. They originate from inter-atomic interactions in atomic gases and the electron-electron interaction in solids. For example, the spin-charge coupling in Mott insulators generates a transverse relaxation term, which induces the decoherence of the wave function\cite{Murakami2022anomalous}. 
Therefore, it is important to clarify the effects of relaxation processes on the HHG, in particular, because in the strongly correlated materials,  as in the Mott insulator $\mathrm{Ca_{2} Ru O_{4}}$, electron-electron interactions and relaxation processes can be large.
In Fig.~\ref{relax}, we show the intensity of the 5$\omega_{0}$-harmonics over the gap width and the strength of the Rabi frequency for three values of the relaxation rate, $\gamma_{l} = 2\gamma_{t} = 0.001,0.03,0.1$.
We note that the range of the gap dependence is limited to the region $\Delta \geq 2.0$ to avoid the effect of a multi-photon resonance at $\Delta = 5\omega_{0}$ so that we can understand the unconventional gap dependence more clearly.

In Fig.~\ref{relax},  we see that while the HHG spectrum is smooth for large relaxation rates, small relaxation rates lead to a fine structure inside the spectrum. However, these figures also show that, regardless of the value of the relaxation rate, the qualitative structure of the HHG spectrum and the unconventional gap dependence remain. The HHG increases for large Rabi frequencies and small gap widths when the gap increases. Thus, we conclude that relaxation processes are not essential for the observation of the unconventional gap dependence in the HHG. However, relaxation processes help observe it by hiding the fine structure.

\subsection{Emission energy dependence of enhancement rate of HHG}
\label{scaling_law}
\begin{figure}[t]
\begin{center}
\includegraphics[width=\linewidth]{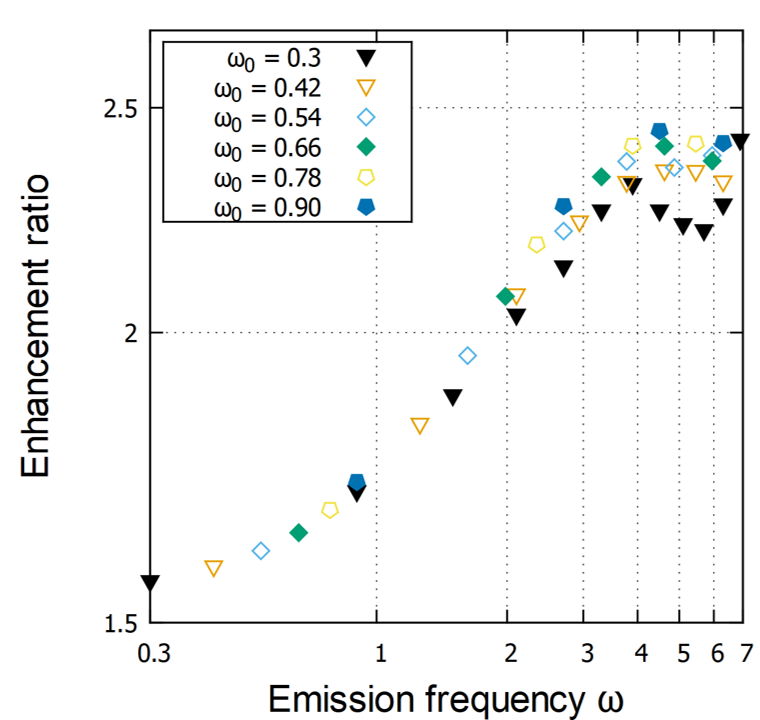}
\end{center}
\caption{Emission energy dependence of enhancement ratio $|p(\omega;\Delta=1.5)|^{2}/|p(\omega;\Delta=1.0)|^{2}$ of HHG. 
The Rabi frequency is $\Omega_{0}=22.7$. Both of the axes are log-scale.
}
\label{enhancement}
\end{figure}
Finally, we show in Fig.~\ref{enhancement} the enhancement ratio for small gap widths, $|p(\omega;\Delta=1.5)|^{2}/|p(\omega;\Delta=1.0)|^{2}$, over the emission frequency for different incidental frequencies, $\omega_0$.
The incidental frequency is varied in the range $\omega_{0}=0.3 \sim 0.9$.
Thus, this figure includes different high harmonics depending on $\omega$ and $\omega_0$.
  This figure shows that the enhancement ratio obeys a universal behavior regardless of the incidental frequency until $\omega \sim 4.0$. It also shows that the enhancement is larger for larger emission energy.
Furthermore, especially around $\omega_{\mathrm{emit}} = 1.0 \sim 3.0$, we can see that the enhancement ratio shows a power law dependence with respect to the emission energy.
This behavior is consistent with the experimental observation of an unconventional scaling law in $\mathrm{Ca_{2} Ru O_{4}}$\cite{uchida2022high}.

\section{Conclusion}
\label{conclusion}
In this paper, we have demonstrated that the HHG is enhanced as the gap width increases if the Rabi frequency is sufficiently large compared to the excitation gap.
This suggests that the unconventional gap dependence, experimentally observed in $\mathrm{Ca_{2}RuO4_{4}}$, originates in significantly strong light-matter coupling.
We note that, in Mott insulators, large nonlinear optical responses have been observed, which also suggests a strong light-matter coupling in Mott insulating systems\cite{kishida2000gigantic,ogasawara2000ultrafast,mizuno2000nonlinear,zhang2001origin}.
Our analysis has revealed that the unconventional gap dependence is related to the saturation and even decrease in polarization towards the center of the pulse when the light-matter coupling is too strong.
Thus, in the regime where the HHG is enhanced by increasing the gap width, the dominant HHG contributions appear at the start and end of the pulse. Increasing the gap width in this regime, we find that the time when the dominant contribution appears moves towards the center of the pulse. The enhancement can thus be understood as an increase in the HHG at the center of the pulse.
On the other hand, a conventional gap dependence is observed when the dominant contribution to the HHG appears only at the center of the peak. 
We note that time-resolved high harmonic spectroscopy, utilizing the pump-probe method\cite{sarantseva2021time}, might be able to observe this signature in  $\mathrm{Ca_{2}RuO4_{4}}$.

Furthermore, we have shown that relaxation processes are unnecessary to induce this gap dependence of the HHG. However, it is easier to observe such a gap dependence in systems with strong relaxation because the relaxation smears fine structures in the dependence of the HHG strength on the gap width.
Finally, we have investigated the enhancement of the emission energy depending on the incidental frequency. We have found a universal behavior demonstrating that the HHG increases faster for larger emission energy, which is also confirmed in semiconductors in the appendix.
This behavior is consistent with observations in $\mathrm{Ca_{2}RuO_{4}}$.

\section*{Acknowledgements}
\label{acknowledgements}
We thank Kento Uchida, Koki Shinada, and Yasushi Shinohara for their insightful discussions.
R.P. is supported by JSPS KAKENHI No.~JP18K03511 and JP23K03300.  This work was supported by JST, the establishment of university fellowships towards the creation of science and technology innovation, and Grant Number JPMJFS2123.
Parts of the numerical simulations in this work have been done using the facilities of the Supercomputer Center at the
Institute for Solid State Physics, the University of Tokyo.

\appendix
\section{Calculations in semiconductors}\label{appenxix_semiconductor}
\begin{figure}[t]
\begin{center}
\includegraphics[width=\linewidth]{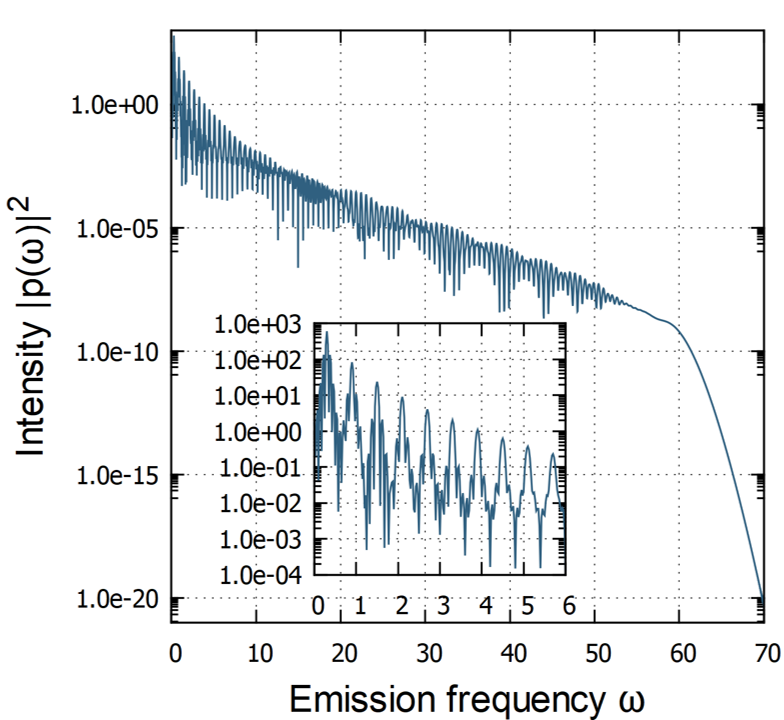}
\end{center}
\caption{A typical example of an HHG spectrum in semiconductors. The inset shows the enlarged view of the spectrum for low frequencies. The parameters are $\Delta=4.85$, $\Omega_{0}=30.25$, $\omega_{0}=0.3$. The vertical axis is log-scale.}
\label{hhg_semiconductor_spectrum}
\end{figure}
\begin{figure}[t]
\begin{center}
\includegraphics[width=\linewidth]{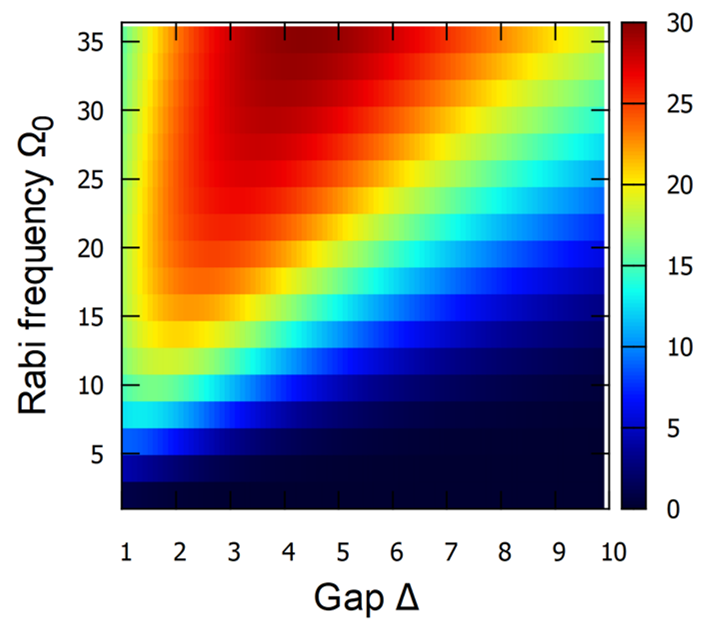}
\end{center}
\caption{Gap and Rabi frequency dependence of the 5th harmonics ($\omega=5\omega_{0}$) in semiconductors. The incidental frequency is $\omega_{0}=0.3$.}
\label{hhg_semiconductor_gap_field}
\end{figure}
In this appendix, we calculate the gap dependence of the HHG in semiconductors and confirm that the results in the main text are not restricted to two-level systems.
Here, we use the semiconductor Bloch equations, which is an extension of the optical Bloch equations\cite{haug2009quantum}.
For simplicity, we assume that our system consists of two bands that are rigid against the incidental electric fields.
The Hamiltonian that  describes the system is
\begin{equation}
    H = \sum_{\bm{k}} (E_{\bm{k}}^{e} \alpha_{\bm{k}}^{\dagger} \alpha_{\bm{k}} + E_{\bm{k}}^{h} \beta_{\bm{-k}}^{\dagger} \beta_{\bm{-k}}) - \sum_{\bm{k}} \Omega(t)(\alpha_{\bm{k}}^{\dagger} \beta_{\bm{-k}}^{\dagger} + h.c.).
\end{equation}
$E_{\bm{k}}^{e}$ and $E_{\bm{k}}^{h}$ are the energy bands of the electrons and holes, $\alpha_{\bm{k}}^{\dagger}$ and $\alpha_{\bm{k}}$ are creation and annihilation operators of the electrons with wavenumber $\bm{k}$, $\beta_{\bm{-k}}^{\dagger}$ and $\beta_{\bm{-k}}$ are creation and annihilation operators of the holes with wavenumber $-\bm{k}$, and $\Omega(t)$ is the light-matter coupling.
Equations of motion for the distribution of the electrons $f_{\bm{k}}^{e} = \langle \alpha_{\bm{k}}^{\dagger} \alpha_{\bm{k}} \rangle$, the  holes $f_{\bm{k}}^{h} = \langle \beta_{\bm{-k}}^{\dagger} \beta_{\bm{-k}} \rangle$, and the  polarization $P_{\bm{k}} = \langle \beta_{\bm{-k}} \alpha_{\bm{k}} \rangle$ are 
\begin{eqnarray}
    \dot{f_{\bm{k}}^{e}} &=& \dot{f_{\bm{k}}^{h}} = -2 \Omega(t) \mathrm{Im} (P_{\bm{k}}) \\
    \dot{P_{\bm{k}}} &=& -i(E_{\bm{k}}^{e} + E_{\bm{k}}^{h})P_{\bm{k}} -i \Omega(t) (f_{\bm{k}}^{e} + f_{\bm{k}}^{h} -1).
\end{eqnarray}
To compare with two-level systems, we define $x_{\bm{k}} = f_{\bm{k}}^{e}+f_{\bm{k}}^{h}-1$, 
$y_{\bm{k}} = P_{\bm{k}}$, and 
$\Delta_{\bm{k}} = E_{\bm{k}}^{e} + E_{\bm{k}}^{h}$. The equations of motion of $x_{\bm{k}}$ and $y_{\bm{k}}$ are
\begin{eqnarray}
\dot{x_{\bm{k}}} &=& -2 i \Omega(t) (y_{\bm{k}}-y_{\bm{k}}^{*}) - \gamma_{l} (x_{\bm{k}}-1)\\
\dot{y_{\bm{k}}} &=& - \left(i \Delta_{\bm{k}} + \gamma_{t} \right)y_{\bm{k}} -i \Omega(t) x_{\bm{k}} .
\end{eqnarray}
Here, we have introduced relaxation terms to the above equation phenomenologically, as done in section \ref{model_and_methods}.
These equations are identical to that of the two-level system with the exemption that the quantities depend on the wavenumber, $\bm{k}$. Thus, the gap also depends on $\bm{k}$.
We define the polarization of the system as the conjugate variable of the external fields (light-matter coupling) as 
\begin{equation}
    p(t) = \left\langle -\frac{\partial H}{\partial \Omega(t)} \right\rangle = 2\sum_{\bm{k}} \mathrm{Re}(y_{\bm{k}})
\end{equation}
For the calculations in this appendix, we consider the dispersion as $\Delta_{\bm{k}} = \delta - t \cos k_{x} \cos k_{y}$ ($ -\pi \leq k_{x},k_{y} \leq \pi$), and an external field $\Omega(t) = \Omega_{0} e^{-\frac{t^{2}}{\tau^{2}}} \cos(\omega_{0} t)$.
In the appendix, the parameters are 
$t = 1.0$, $\gamma_{l}=2.0$, $\gamma_{t}=1.0$, $\tau=8.0\pi$, and we vary $\delta$, $\Omega_{0}$, and $\omega_{0}$.
The gap of the system is defined as $\Delta = \delta -t$.

\begin{figure}[t]
\begin{center}
\includegraphics[width=\linewidth]{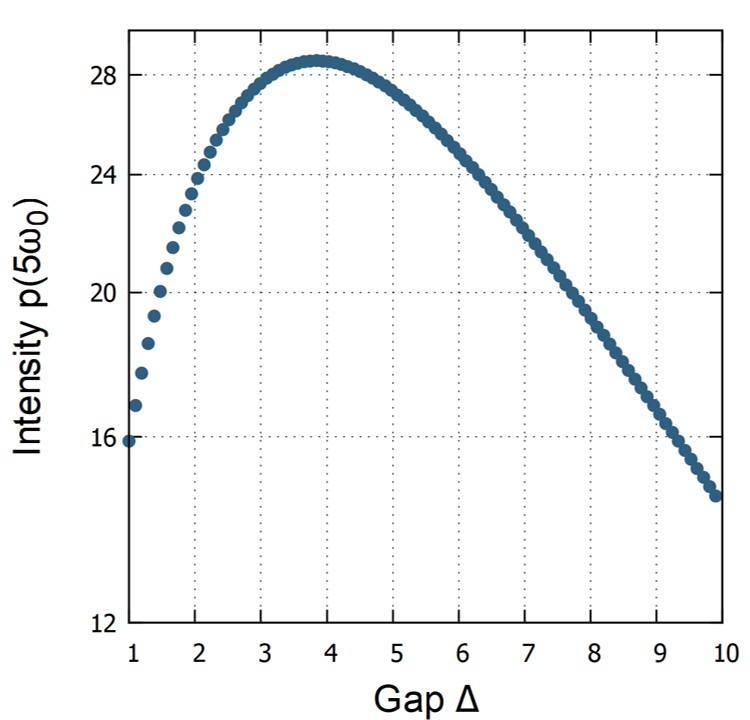}
\end{center}
\caption{Gap dependence of the 5th harmonics in semiconductors for $\Omega_{0}=30.25$. The incidental frequency is $\omega_{0}=0.3$. The vertical axis is log-scale.}
\label{hhg_semiconductor_gap_dep}
\end{figure}
First, we show a typical HHG spectrum calculated for a semiconductor in Fig.~\ref{hhg_semiconductor_spectrum}. The spectrum shows a plateau and cutoff energy similar to two-level systems.
Figure~\ref{hhg_semiconductor_gap_field} shows the intensity of the $5\omega_{0}$-harmonics for various gap widths and Rabi frequencies. It becomes clear that for sufficiently large values of the Rabi frequency compared to gap width, the intensity increases as the gap width is increased. The intensity of the $5\omega_{0}$-harmonics takes a maximum around $\Omega_{0}/\Delta \sim 7$.
To confirm this behavior, we also show the gap dependence of the HHG at $\Omega_{0}=30.25$ in Fig.~\ref{hhg_semiconductor_gap_dep}. The figure shows that the intensity increases until the gap width reaches some threshold and then decreases exponentially.

\begin{figure}[t]
\begin{center}
\includegraphics[width=\linewidth]{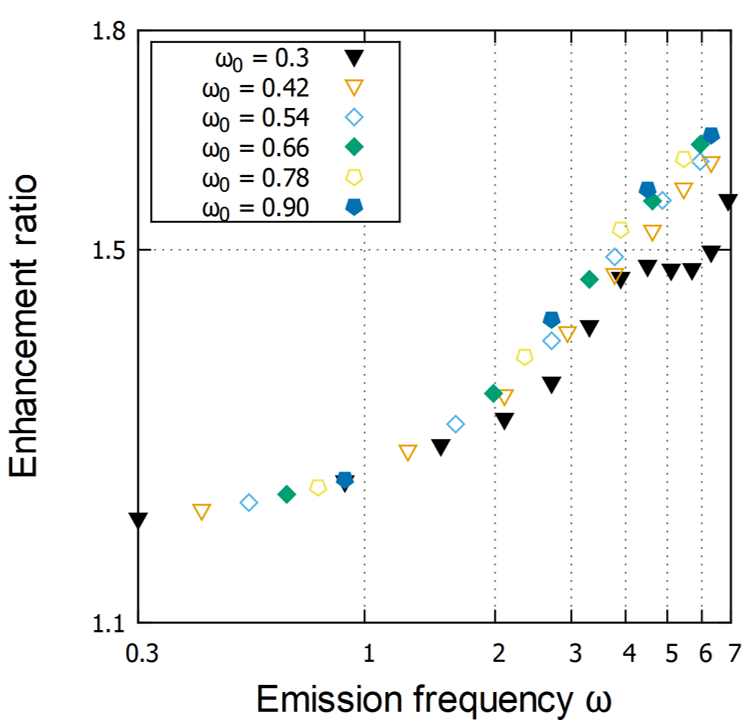}
\end{center}
\caption{Emission energy dependence of the enhancement ratio $|p(\omega;\Delta=1.5)|^{2}/|p(\omega;\Delta=1.0)|^{2}$ of the HHG in semiconductors. The Rabi frequency is $\Omega_{0}=30.25$. Both of the axes are log-scale.}
\label{hhg_semiconductor_enhance}
\end{figure}
Next, we show the enhancement ratio over the emission frequency ($|p(\omega;\Delta=1.5)|^{2}/|p(\omega;\Delta=1.0)|^{2}$) for various incidental frequencies in Fig.~\ref{hhg_semiconductor_enhance}.
The incidental frequency is varied in the range $\omega_{0} = 0.3\sim 0.9$.
The figure reveals that the enhancement ratio increases for increasing emission frequencies, demonstrating a universal scaling law that does not depend on the incidental frequency.

Thus, we conclude that when the Rabi frequency is sufficiently large compared to the gap width, HHG in semiconductors also shows an unconventional gap dependence similar to the two-level systems studied in the main text.
This clearly shows that the origin of the unconventional gap dependence in the main text cannot be traced back to some particularity of two-level systems.

\bibliography{bunken}

\begin{thebibliography}{59}%
\makeatletter
\providecommand \@ifxundefined [1]{%
 \@ifx{#1\undefined}
}%
\providecommand \@ifnum [1]{%
 \ifnum #1\expandafter \@firstoftwo
 \else \expandafter \@secondoftwo
 \fi
}%
\providecommand \@ifx [1]{%
 \ifx #1\expandafter \@firstoftwo
 \else \expandafter \@secondoftwo
 \fi
}%
\providecommand \natexlab [1]{#1}%
\providecommand \enquote  [1]{``#1''}%
\providecommand \bibnamefont  [1]{#1}%
\providecommand \bibfnamefont [1]{#1}%
\providecommand \citenamefont [1]{#1}%
\providecommand \href@noop [0]{\@secondoftwo}%
\providecommand \href [0]{\begingroup \@sanitize@url \@href}%
\providecommand \@href[1]{\@@startlink{#1}\@@href}%
\providecommand \@@href[1]{\endgroup#1\@@endlink}%
\providecommand \@sanitize@url [0]{\catcode `\\12\catcode `\$12\catcode
  `\&12\catcode `\#12\catcode `\^12\catcode `\_12\catcode `\%12\relax}%
\providecommand \@@startlink[1]{}%
\providecommand \@@endlink[0]{}%
\providecommand \url  [0]{\begingroup\@sanitize@url \@url }%
\providecommand \@url [1]{\endgroup\@href {#1}{\urlprefix }}%
\providecommand \urlprefix  [0]{URL }%
\providecommand \Eprint [0]{\href }%
\providecommand \doibase [0]{http://dx.doi.org/}%
\providecommand \selectlanguage [0]{\@gobble}%
\providecommand \bibinfo  [0]{\@secondoftwo}%
\providecommand \bibfield  [0]{\@secondoftwo}%
\providecommand \translation [1]{[#1]}%
\providecommand \BibitemOpen [0]{}%
\providecommand \bibitemStop [0]{}%
\providecommand \bibitemNoStop [0]{.\EOS\space}%
\providecommand \EOS [0]{\spacefactor3000\relax}%
\providecommand \BibitemShut  [1]{\csname bibitem#1\endcsname}%
\let\auto@bib@innerbib\@empty
%</preamble>
\bibitem [{\citenamefont {McPherson}\ \emph {et~al.}(1987)\citenamefont
  {McPherson}, \citenamefont {Gibson}, \citenamefont {Jara}, \citenamefont
  {Johann}, \citenamefont {Luk}, \citenamefont {McIntyre}, \citenamefont
  {Boyer},\ and\ \citenamefont {Rhodes}}]{mcpherson1987studies}%
  \BibitemOpen
  \bibfield  {author} {\bibinfo {author} {\bibfnamefont {A}~\bibnamefont
  {McPherson}}, \bibinfo {author} {\bibfnamefont {G}~\bibnamefont {Gibson}},
  \bibinfo {author} {\bibfnamefont {H}~\bibnamefont {Jara}}, \bibinfo {author}
  {\bibfnamefont {U}~\bibnamefont {Johann}}, \bibinfo {author} {\bibfnamefont
  {Ting~S}\ \bibnamefont {Luk}}, \bibinfo {author} {\bibfnamefont
  {IA}~\bibnamefont {McIntyre}}, \bibinfo {author} {\bibfnamefont {Keith}\
  \bibnamefont {Boyer}}, \ and\ \bibinfo {author} {\bibfnamefont {Charles~K}\
  \bibnamefont {Rhodes}},\ }\bibfield  {title} {\enquote {\bibinfo {title}
  {Studies of multiphoton production of vacuum-ultraviolet radiation in the
  rare gases},}\ }\href@noop {} {\bibfield  {journal} {\bibinfo  {journal}
  {JOSA B}\ }\textbf {\bibinfo {volume} {4}},\ \bibinfo {pages} {595--601}
  (\bibinfo {year} {1987})}\BibitemShut {NoStop}%
\bibitem [{\citenamefont {Ferray}\ \emph {et~al.}(1988)\citenamefont {Ferray},
  \citenamefont {L'Huillier}, \citenamefont {Li}, \citenamefont {Lompre},
  \citenamefont {Mainfray},\ and\ \citenamefont {Manus}}]{ferray1988multiple}%
  \BibitemOpen
  \bibfield  {author} {\bibinfo {author} {\bibfnamefont {M}~\bibnamefont
  {Ferray}}, \bibinfo {author} {\bibfnamefont {Anne}\ \bibnamefont
  {L'Huillier}}, \bibinfo {author} {\bibfnamefont {XF}~\bibnamefont {Li}},
  \bibinfo {author} {\bibfnamefont {LA}~\bibnamefont {Lompre}}, \bibinfo
  {author} {\bibfnamefont {G}~\bibnamefont {Mainfray}}, \ and\ \bibinfo
  {author} {\bibfnamefont {C}~\bibnamefont {Manus}},\ }\bibfield  {title}
  {\enquote {\bibinfo {title} {Multiple-harmonic conversion of 1064 nm
  radiation in rare gases},}\ }\href@noop {} {\bibfield  {journal} {\bibinfo
  {journal} {Journal of Physics B: Atomic, Molecular and Optical Physics}\
  }\textbf {\bibinfo {volume} {21}},\ \bibinfo {pages} {L31} (\bibinfo {year}
  {1988})}\BibitemShut {NoStop}%
\bibitem [{\citenamefont {Krausz}\ and\ \citenamefont
  {Ivanov}(2009)}]{krausz2009attosecond}%
  \BibitemOpen
  \bibfield  {author} {\bibinfo {author} {\bibfnamefont {Ferenc}\ \bibnamefont
  {Krausz}}\ and\ \bibinfo {author} {\bibfnamefont {Misha}\ \bibnamefont
  {Ivanov}},\ }\bibfield  {title} {\enquote {\bibinfo {title} {Attosecond
  physics},}\ }\href@noop {} {\bibfield  {journal} {\bibinfo  {journal}
  {Reviews of modern physics}\ }\textbf {\bibinfo {volume} {81}},\ \bibinfo
  {pages} {163} (\bibinfo {year} {2009})}\BibitemShut {NoStop}%
\bibitem [{\citenamefont {Ghimire}\ \emph {et~al.}(2011)\citenamefont
  {Ghimire}, \citenamefont {DiChiara}, \citenamefont {Sistrunk}, \citenamefont
  {Agostini}, \citenamefont {DiMauro},\ and\ \citenamefont
  {Reis}}]{ghimire2011observation}%
  \BibitemOpen
  \bibfield  {author} {\bibinfo {author} {\bibfnamefont {Shambhu}\ \bibnamefont
  {Ghimire}}, \bibinfo {author} {\bibfnamefont {Anthony~D}\ \bibnamefont
  {DiChiara}}, \bibinfo {author} {\bibfnamefont {Emily}\ \bibnamefont
  {Sistrunk}}, \bibinfo {author} {\bibfnamefont {Pierre}\ \bibnamefont
  {Agostini}}, \bibinfo {author} {\bibfnamefont {Louis~F}\ \bibnamefont
  {DiMauro}}, \ and\ \bibinfo {author} {\bibfnamefont {David~A}\ \bibnamefont
  {Reis}},\ }\bibfield  {title} {\enquote {\bibinfo {title} {Observation of
  high-order harmonic generation in a bulk crystal},}\ }\href@noop {}
  {\bibfield  {journal} {\bibinfo  {journal} {Nature physics}\ }\textbf
  {\bibinfo {volume} {7}},\ \bibinfo {pages} {138--141} (\bibinfo {year}
  {2011})}\BibitemShut {NoStop}%
\bibitem [{\citenamefont {Ghimire}\ and\ \citenamefont
  {Reis}(2019)}]{ghimire2019high}%
  \BibitemOpen
  \bibfield  {author} {\bibinfo {author} {\bibfnamefont {Shambhu}\ \bibnamefont
  {Ghimire}}\ and\ \bibinfo {author} {\bibfnamefont {David~A}\ \bibnamefont
  {Reis}},\ }\bibfield  {title} {\enquote {\bibinfo {title} {High-harmonic
  generation from solids},}\ }\href@noop {} {\bibfield  {journal} {\bibinfo
  {journal} {Nature physics}\ }\textbf {\bibinfo {volume} {15}},\ \bibinfo
  {pages} {10--16} (\bibinfo {year} {2019})}\BibitemShut {NoStop}%
\bibitem [{\citenamefont {Itatani}\ \emph {et~al.}(2004)\citenamefont
  {Itatani}, \citenamefont {Levesque}, \citenamefont {Zeidler}, \citenamefont
  {Niikura}, \citenamefont {P{\'e}pin}, \citenamefont {Kieffer}, \citenamefont
  {Corkum},\ and\ \citenamefont {Villeneuve}}]{itatani2004tomographic}%
  \BibitemOpen
  \bibfield  {author} {\bibinfo {author} {\bibfnamefont {Jiro}\ \bibnamefont
  {Itatani}}, \bibinfo {author} {\bibfnamefont {J{\'e}r{\^o}me}\ \bibnamefont
  {Levesque}}, \bibinfo {author} {\bibfnamefont {Dirk}\ \bibnamefont
  {Zeidler}}, \bibinfo {author} {\bibfnamefont {Hiromichi}\ \bibnamefont
  {Niikura}}, \bibinfo {author} {\bibfnamefont {Henri}\ \bibnamefont
  {P{\'e}pin}}, \bibinfo {author} {\bibfnamefont {Jean-Claude}\ \bibnamefont
  {Kieffer}}, \bibinfo {author} {\bibfnamefont {Paul~B}\ \bibnamefont
  {Corkum}}, \ and\ \bibinfo {author} {\bibfnamefont {David~M}\ \bibnamefont
  {Villeneuve}},\ }\bibfield  {title} {\enquote {\bibinfo {title} {Tomographic
  imaging of molecular orbitals},}\ }\href@noop {} {\bibfield  {journal}
  {\bibinfo  {journal} {Nature}\ }\textbf {\bibinfo {volume} {432}},\ \bibinfo
  {pages} {867--871} (\bibinfo {year} {2004})}\BibitemShut {NoStop}%
\bibitem [{\citenamefont {Wagner}\ \emph {et~al.}(2006)\citenamefont {Wagner},
  \citenamefont {W{\"u}est}, \citenamefont {Christov}, \citenamefont
  {Popmintchev}, \citenamefont {Zhou}, \citenamefont {Murnane},\ and\
  \citenamefont {Kapteyn}}]{wagner2006monitoring}%
  \BibitemOpen
  \bibfield  {author} {\bibinfo {author} {\bibfnamefont {Nicholas~L}\
  \bibnamefont {Wagner}}, \bibinfo {author} {\bibfnamefont {Andrea}\
  \bibnamefont {W{\"u}est}}, \bibinfo {author} {\bibfnamefont {Ivan~P}\
  \bibnamefont {Christov}}, \bibinfo {author} {\bibfnamefont {Tenio}\
  \bibnamefont {Popmintchev}}, \bibinfo {author} {\bibfnamefont {Xibin}\
  \bibnamefont {Zhou}}, \bibinfo {author} {\bibfnamefont {Margaret~M}\
  \bibnamefont {Murnane}}, \ and\ \bibinfo {author} {\bibfnamefont {Henry~C}\
  \bibnamefont {Kapteyn}},\ }\bibfield  {title} {\enquote {\bibinfo {title}
  {Monitoring molecular dynamics using coherent electrons from high harmonic
  generation},}\ }\href@noop {} {\bibfield  {journal} {\bibinfo  {journal}
  {Proceedings of the National Academy of Sciences}\ }\textbf {\bibinfo
  {volume} {103}},\ \bibinfo {pages} {13279--13285} (\bibinfo {year}
  {2006})}\BibitemShut {NoStop}%
\bibitem [{\citenamefont {Baker}\ \emph {et~al.}(2006)\citenamefont {Baker},
  \citenamefont {Robinson}, \citenamefont {Haworth}, \citenamefont {Teng},
  \citenamefont {Smith}, \citenamefont {Chirilă}, \citenamefont {Lein},
  \citenamefont {Tisch},\ and\ \citenamefont {Marangos}}]{baker2006probing}%
  \BibitemOpen
  \bibfield  {author} {\bibinfo {author} {\bibfnamefont {Sarah}\ \bibnamefont
  {Baker}}, \bibinfo {author} {\bibfnamefont {Joseph~S}\ \bibnamefont
  {Robinson}}, \bibinfo {author} {\bibfnamefont {CA}~\bibnamefont {Haworth}},
  \bibinfo {author} {\bibfnamefont {H}~\bibnamefont {Teng}}, \bibinfo {author}
  {\bibfnamefont {RA}~\bibnamefont {Smith}}, \bibinfo {author} {\bibfnamefont
  {Ciprian~C}\ \bibnamefont {Chirilă}}, \bibinfo {author} {\bibfnamefont
  {Manfred}\ \bibnamefont {Lein}}, \bibinfo {author} {\bibfnamefont {JWG}\
  \bibnamefont {Tisch}}, \ and\ \bibinfo {author} {\bibfnamefont {Jonathan~P}\
  \bibnamefont {Marangos}},\ }\bibfield  {title} {\enquote {\bibinfo {title}
  {Probing proton dynamics in molecules on an attosecond time scale},}\
  }\href@noop {} {\bibfield  {journal} {\bibinfo  {journal} {Science}\ }\textbf
  {\bibinfo {volume} {312}},\ \bibinfo {pages} {424--427} (\bibinfo {year}
  {2006})}\BibitemShut {NoStop}%
\bibitem [{\citenamefont {Patchkovskii}(2009)}]{patchkovskii2009nuclear}%
  \BibitemOpen
  \bibfield  {author} {\bibinfo {author} {\bibfnamefont {Serguei}\ \bibnamefont
  {Patchkovskii}},\ }\bibfield  {title} {\enquote {\bibinfo {title} {Nuclear
  dynamics in polyatomic molecules and high-order harmonic generation},}\
  }\href@noop {} {\bibfield  {journal} {\bibinfo  {journal} {Physical Review
  Letters}\ }\textbf {\bibinfo {volume} {102}},\ \bibinfo {pages} {253602}
  (\bibinfo {year} {2009})}\BibitemShut {NoStop}%
\bibitem [{\citenamefont {W{\"o}rner}\ \emph {et~al.}(2011)\citenamefont
  {W{\"o}rner}, \citenamefont {Bertrand}, \citenamefont {Fabre}, \citenamefont
  {Higuet}, \citenamefont {Ruf}, \citenamefont {Dubrouil}, \citenamefont
  {Patchkovskii}, \citenamefont {Spanner}, \citenamefont {Mairesse},
  \citenamefont {Blanchet} \emph {et~al.}}]{worner2011conical}%
  \BibitemOpen
  \bibfield  {author} {\bibinfo {author} {\bibfnamefont {Hans~Jakob}\
  \bibnamefont {W{\"o}rner}}, \bibinfo {author} {\bibfnamefont {Julien~B}\
  \bibnamefont {Bertrand}}, \bibinfo {author} {\bibfnamefont {Baptiste}\
  \bibnamefont {Fabre}}, \bibinfo {author} {\bibfnamefont {Julien}\
  \bibnamefont {Higuet}}, \bibinfo {author} {\bibfnamefont {Hartmut}\
  \bibnamefont {Ruf}}, \bibinfo {author} {\bibfnamefont {Antoine}\ \bibnamefont
  {Dubrouil}}, \bibinfo {author} {\bibfnamefont {Serguei}\ \bibnamefont
  {Patchkovskii}}, \bibinfo {author} {\bibfnamefont {Michael}\ \bibnamefont
  {Spanner}}, \bibinfo {author} {\bibfnamefont {Yann}\ \bibnamefont
  {Mairesse}}, \bibinfo {author} {\bibfnamefont {Val{\'e}rie}\ \bibnamefont
  {Blanchet}},  \emph {et~al.},\ }\bibfield  {title} {\enquote {\bibinfo
  {title} {Conical intersection dynamics in no2 probed by homodyne
  high-harmonic spectroscopy},}\ }\href@noop {} {\bibfield  {journal} {\bibinfo
   {journal} {Science}\ }\textbf {\bibinfo {volume} {334}},\ \bibinfo {pages}
  {208--212} (\bibinfo {year} {2011})}\BibitemShut {NoStop}%
\bibitem [{\citenamefont {Le}\ \emph {et~al.}(2012)\citenamefont {Le},
  \citenamefont {Morishita}, \citenamefont {Lucchese},\ and\ \citenamefont
  {Lin}}]{le2012theory}%
  \BibitemOpen
  \bibfield  {author} {\bibinfo {author} {\bibfnamefont {Anh-Thu}\ \bibnamefont
  {Le}}, \bibinfo {author} {\bibfnamefont {Toru}\ \bibnamefont {Morishita}},
  \bibinfo {author} {\bibfnamefont {RR}~\bibnamefont {Lucchese}}, \ and\
  \bibinfo {author} {\bibfnamefont {Chii~Dong}\ \bibnamefont {Lin}},\
  }\bibfield  {title} {\enquote {\bibinfo {title} {Theory of high harmonic
  generation for probing time-resolved large-amplitude molecular vibrations
  with ultrashort intense lasers},}\ }\href@noop {} {\bibfield  {journal}
  {\bibinfo  {journal} {Physical Review Letters}\ }\textbf {\bibinfo {volume}
  {109}},\ \bibinfo {pages} {203004} (\bibinfo {year} {2012})}\BibitemShut
  {NoStop}%
\bibitem [{\citenamefont {Vampa}\ \emph
  {et~al.}(2015{\natexlab{a}})\citenamefont {Vampa}, \citenamefont {Hammond},
  \citenamefont {Thir{\'e}}, \citenamefont {Schmidt}, \citenamefont
  {L{\'e}gar{\'e}}, \citenamefont {McDonald}, \citenamefont {Brabec},
  \citenamefont {Klug},\ and\ \citenamefont {Corkum}}]{vampa2015all}%
  \BibitemOpen
  \bibfield  {author} {\bibinfo {author} {\bibfnamefont {Giulio}\ \bibnamefont
  {Vampa}}, \bibinfo {author} {\bibfnamefont {TJ}~\bibnamefont {Hammond}},
  \bibinfo {author} {\bibfnamefont {Nicolas}\ \bibnamefont {Thir{\'e}}},
  \bibinfo {author} {\bibfnamefont {BE}~\bibnamefont {Schmidt}}, \bibinfo
  {author} {\bibfnamefont {Fran{\c{c}}ois}\ \bibnamefont {L{\'e}gar{\'e}}},
  \bibinfo {author} {\bibfnamefont {CR}~\bibnamefont {McDonald}}, \bibinfo
  {author} {\bibfnamefont {Thomas}\ \bibnamefont {Brabec}}, \bibinfo {author}
  {\bibfnamefont {DD}~\bibnamefont {Klug}}, \ and\ \bibinfo {author}
  {\bibfnamefont {PB}~\bibnamefont {Corkum}},\ }\bibfield  {title} {\enquote
  {\bibinfo {title} {All-optical reconstruction of crystal band structure},}\
  }\href@noop {} {\bibfield  {journal} {\bibinfo  {journal} {Physical review
  letters}\ }\textbf {\bibinfo {volume} {115}},\ \bibinfo {pages} {193603}
  (\bibinfo {year} {2015}{\natexlab{a}})}\BibitemShut {NoStop}%
\bibitem [{\citenamefont {Luu}\ \emph {et~al.}(2015)\citenamefont {Luu},
  \citenamefont {Garg}, \citenamefont {Kruchinin}, \citenamefont {Moulet},
  \citenamefont {Hassan},\ and\ \citenamefont {Goulielmakis}}]{luu2015extreme}%
  \BibitemOpen
  \bibfield  {author} {\bibinfo {author} {\bibfnamefont {Tran~Trung}\
  \bibnamefont {Luu}}, \bibinfo {author} {\bibfnamefont {M}~\bibnamefont
  {Garg}}, \bibinfo {author} {\bibfnamefont {S~Yu}\ \bibnamefont {Kruchinin}},
  \bibinfo {author} {\bibfnamefont {Antoine}\ \bibnamefont {Moulet}}, \bibinfo
  {author} {\bibfnamefont {M~Th}\ \bibnamefont {Hassan}}, \ and\ \bibinfo
  {author} {\bibfnamefont {Eleftherios}\ \bibnamefont {Goulielmakis}},\
  }\bibfield  {title} {\enquote {\bibinfo {title} {Extreme ultraviolet
  high-harmonic spectroscopy of solids},}\ }\href@noop {} {\bibfield  {journal}
  {\bibinfo  {journal} {Nature}\ }\textbf {\bibinfo {volume} {521}},\ \bibinfo
  {pages} {498--502} (\bibinfo {year} {2015})}\BibitemShut {NoStop}%
\bibitem [{\citenamefont {Hohenleutner}\ \emph {et~al.}(2015)\citenamefont
  {Hohenleutner}, \citenamefont {Langer}, \citenamefont {Schubert},
  \citenamefont {Knorr}, \citenamefont {Huttner}, \citenamefont {Koch},
  \citenamefont {Kira},\ and\ \citenamefont {Huber}}]{hohenleutner2015real}%
  \BibitemOpen
  \bibfield  {author} {\bibinfo {author} {\bibfnamefont {Matthias}\
  \bibnamefont {Hohenleutner}}, \bibinfo {author} {\bibfnamefont {Fabian}\
  \bibnamefont {Langer}}, \bibinfo {author} {\bibfnamefont {Olaf}\ \bibnamefont
  {Schubert}}, \bibinfo {author} {\bibfnamefont {Matthias}\ \bibnamefont
  {Knorr}}, \bibinfo {author} {\bibfnamefont {U}~\bibnamefont {Huttner}},
  \bibinfo {author} {\bibfnamefont {Stephan~W}\ \bibnamefont {Koch}}, \bibinfo
  {author} {\bibfnamefont {M}~\bibnamefont {Kira}}, \ and\ \bibinfo {author}
  {\bibfnamefont {Rupert}\ \bibnamefont {Huber}},\ }\bibfield  {title}
  {\enquote {\bibinfo {title} {Real-time observation of interfering crystal
  electrons in high-harmonic generation},}\ }\href@noop {} {\bibfield
  {journal} {\bibinfo  {journal} {Nature}\ }\textbf {\bibinfo {volume} {523}},\
  \bibinfo {pages} {572--575} (\bibinfo {year} {2015})}\BibitemShut {NoStop}%
\bibitem [{\citenamefont {Lein}(2005)}]{lein2005attosecond}%
  \BibitemOpen
  \bibfield  {author} {\bibinfo {author} {\bibfnamefont {Manfred}\ \bibnamefont
  {Lein}},\ }\bibfield  {title} {\enquote {\bibinfo {title} {Attosecond probing
  of vibrational dynamics with high-harmonic generation},}\ }\href@noop {}
  {\bibfield  {journal} {\bibinfo  {journal} {Physical review letters}\
  }\textbf {\bibinfo {volume} {94}},\ \bibinfo {pages} {053004} (\bibinfo
  {year} {2005})}\BibitemShut {NoStop}%
\bibitem [{\citenamefont {You}\ \emph {et~al.}(2017)\citenamefont {You},
  \citenamefont {Reis},\ and\ \citenamefont {Ghimire}}]{you2017anisotropic}%
  \BibitemOpen
  \bibfield  {author} {\bibinfo {author} {\bibfnamefont {Yong~Sing}\
  \bibnamefont {You}}, \bibinfo {author} {\bibfnamefont {David~A}\ \bibnamefont
  {Reis}}, \ and\ \bibinfo {author} {\bibfnamefont {Shambhu}\ \bibnamefont
  {Ghimire}},\ }\bibfield  {title} {\enquote {\bibinfo {title} {Anisotropic
  high-harmonic generation in bulk crystals},}\ }\href@noop {} {\bibfield
  {journal} {\bibinfo  {journal} {Nature physics}\ }\textbf {\bibinfo {volume}
  {13}},\ \bibinfo {pages} {345--349} (\bibinfo {year} {2017})}\BibitemShut
  {NoStop}%
\bibitem [{\citenamefont {Kaneshima}\ \emph {et~al.}(2018)\citenamefont
  {Kaneshima}, \citenamefont {Shinohara}, \citenamefont {Takeuchi},
  \citenamefont {Ishii}, \citenamefont {Imasaka}, \citenamefont {Kaji},
  \citenamefont {Ashihara}, \citenamefont {Ishikawa},\ and\ \citenamefont
  {Itatani}}]{kaneshima2018polarization}%
  \BibitemOpen
  \bibfield  {author} {\bibinfo {author} {\bibfnamefont {Keisuke}\ \bibnamefont
  {Kaneshima}}, \bibinfo {author} {\bibfnamefont {Yasushi}\ \bibnamefont
  {Shinohara}}, \bibinfo {author} {\bibfnamefont {Kengo}\ \bibnamefont
  {Takeuchi}}, \bibinfo {author} {\bibfnamefont {Nobuhisa}\ \bibnamefont
  {Ishii}}, \bibinfo {author} {\bibfnamefont {Kotaro}\ \bibnamefont {Imasaka}},
  \bibinfo {author} {\bibfnamefont {Tomohiro}\ \bibnamefont {Kaji}}, \bibinfo
  {author} {\bibfnamefont {Satoshi}\ \bibnamefont {Ashihara}}, \bibinfo
  {author} {\bibfnamefont {Kenichi~L}\ \bibnamefont {Ishikawa}}, \ and\
  \bibinfo {author} {\bibfnamefont {Jiro}\ \bibnamefont {Itatani}},\ }\bibfield
   {title} {\enquote {\bibinfo {title} {Polarization-resolved study of high
  harmonics from bulk semiconductors},}\ }\href@noop {} {\bibfield  {journal}
  {\bibinfo  {journal} {Physical review letters}\ }\textbf {\bibinfo {volume}
  {120}},\ \bibinfo {pages} {243903} (\bibinfo {year} {2018})}\BibitemShut
  {NoStop}%
\bibitem [{\citenamefont {Luu}\ and\ \citenamefont
  {W{\"o}rner}(2018)}]{luu2018measurement}%
  \BibitemOpen
  \bibfield  {author} {\bibinfo {author} {\bibfnamefont {Tran~Trung}\
  \bibnamefont {Luu}}\ and\ \bibinfo {author} {\bibfnamefont {Hans~Jakob}\
  \bibnamefont {W{\"o}rner}},\ }\bibfield  {title} {\enquote {\bibinfo {title}
  {Measurement of the berry curvature of solids using high-harmonic
  spectroscopy},}\ }\href@noop {} {\bibfield  {journal} {\bibinfo  {journal}
  {Nature communications}\ }\textbf {\bibinfo {volume} {9}},\ \bibinfo {pages}
  {916} (\bibinfo {year} {2018})}\BibitemShut {NoStop}%
\bibitem [{\citenamefont {Lakhotia}\ \emph {et~al.}(2020)\citenamefont
  {Lakhotia}, \citenamefont {Kim}, \citenamefont {Zhan}, \citenamefont {Hu},
  \citenamefont {Meng},\ and\ \citenamefont
  {Goulielmakis}}]{lakhotia2020laser}%
  \BibitemOpen
  \bibfield  {author} {\bibinfo {author} {\bibfnamefont {H}~\bibnamefont
  {Lakhotia}}, \bibinfo {author} {\bibfnamefont {HY}~\bibnamefont {Kim}},
  \bibinfo {author} {\bibfnamefont {Minjie}\ \bibnamefont {Zhan}}, \bibinfo
  {author} {\bibfnamefont {S}~\bibnamefont {Hu}}, \bibinfo {author}
  {\bibfnamefont {S}~\bibnamefont {Meng}}, \ and\ \bibinfo {author}
  {\bibfnamefont {Eleftherios}\ \bibnamefont {Goulielmakis}},\ }\bibfield
  {title} {\enquote {\bibinfo {title} {Laser picoscopy of valence electrons in
  solids},}\ }\href@noop {} {\bibfield  {journal} {\bibinfo  {journal}
  {Nature}\ }\textbf {\bibinfo {volume} {583}},\ \bibinfo {pages} {55--59}
  (\bibinfo {year} {2020})}\BibitemShut {NoStop}%
\bibitem [{\citenamefont {Uchida}\ \emph {et~al.}(2021)\citenamefont {Uchida},
  \citenamefont {Pareek}, \citenamefont {Nagai}, \citenamefont {Dani},\ and\
  \citenamefont {Tanaka}}]{uchida2021visualization}%
  \BibitemOpen
  \bibfield  {author} {\bibinfo {author} {\bibfnamefont {K}~\bibnamefont
  {Uchida}}, \bibinfo {author} {\bibfnamefont {V}~\bibnamefont {Pareek}},
  \bibinfo {author} {\bibfnamefont {K}~\bibnamefont {Nagai}}, \bibinfo {author}
  {\bibfnamefont {KM}~\bibnamefont {Dani}}, \ and\ \bibinfo {author}
  {\bibfnamefont {K}~\bibnamefont {Tanaka}},\ }\bibfield  {title} {\enquote
  {\bibinfo {title} {Visualization of two-dimensional transition dipole moment
  texture in momentum space using high-harmonic generation spectroscopy},}\
  }\href@noop {} {\bibfield  {journal} {\bibinfo  {journal} {Physical Review
  B}\ }\textbf {\bibinfo {volume} {103}},\ \bibinfo {pages} {L161406} (\bibinfo
  {year} {2021})}\BibitemShut {NoStop}%
\bibitem [{\citenamefont {Bionta}\ \emph {et~al.}(2021)\citenamefont {Bionta},
  \citenamefont {Haddad}, \citenamefont {Leblanc}, \citenamefont {Gruson},
  \citenamefont {Lassonde}, \citenamefont {Ibrahim}, \citenamefont {Chaillou},
  \citenamefont {{\'E}mond}, \citenamefont {Otto}, \citenamefont
  {Jim{\'e}nez-Gal{\'a}n} \emph {et~al.}}]{bionta2021tracking}%
  \BibitemOpen
  \bibfield  {author} {\bibinfo {author} {\bibfnamefont {Mina~R}\ \bibnamefont
  {Bionta}}, \bibinfo {author} {\bibfnamefont {Elissa}\ \bibnamefont {Haddad}},
  \bibinfo {author} {\bibfnamefont {Adrien}\ \bibnamefont {Leblanc}}, \bibinfo
  {author} {\bibfnamefont {Vincent}\ \bibnamefont {Gruson}}, \bibinfo {author}
  {\bibfnamefont {Philippe}\ \bibnamefont {Lassonde}}, \bibinfo {author}
  {\bibfnamefont {Heide}\ \bibnamefont {Ibrahim}}, \bibinfo {author}
  {\bibfnamefont {J{\'e}r{\'e}mie}\ \bibnamefont {Chaillou}}, \bibinfo {author}
  {\bibfnamefont {Nicolas}\ \bibnamefont {{\'E}mond}}, \bibinfo {author}
  {\bibfnamefont {Martin~R}\ \bibnamefont {Otto}}, \bibinfo {author}
  {\bibfnamefont {{\'A}lvaro}\ \bibnamefont {Jim{\'e}nez-Gal{\'a}n}},  \emph
  {et~al.},\ }\bibfield  {title} {\enquote {\bibinfo {title} {Tracking
  ultrafast solid-state dynamics using high harmonic spectroscopy},}\
  }\href@noop {} {\bibfield  {journal} {\bibinfo  {journal} {Physical Review
  Research}\ }\textbf {\bibinfo {volume} {3}},\ \bibinfo {pages} {023250}
  (\bibinfo {year} {2021})}\BibitemShut {NoStop}%
\bibitem [{\citenamefont {Heide}\ \emph {et~al.}(2022)\citenamefont {Heide},
  \citenamefont {Kobayashi}, \citenamefont {Baykusheva}, \citenamefont {Jain},
  \citenamefont {Sobota}, \citenamefont {Hashimoto}, \citenamefont {Kirchmann},
  \citenamefont {Oh}, \citenamefont {Heinz}, \citenamefont {Reis} \emph
  {et~al.}}]{heide2022probing}%
  \BibitemOpen
  \bibfield  {author} {\bibinfo {author} {\bibfnamefont {Christian}\
  \bibnamefont {Heide}}, \bibinfo {author} {\bibfnamefont {Yuki}\ \bibnamefont
  {Kobayashi}}, \bibinfo {author} {\bibfnamefont {Denitsa~R}\ \bibnamefont
  {Baykusheva}}, \bibinfo {author} {\bibfnamefont {Deepti}\ \bibnamefont
  {Jain}}, \bibinfo {author} {\bibfnamefont {Jonathan~A}\ \bibnamefont
  {Sobota}}, \bibinfo {author} {\bibfnamefont {Makoto}\ \bibnamefont
  {Hashimoto}}, \bibinfo {author} {\bibfnamefont {Patrick~S}\ \bibnamefont
  {Kirchmann}}, \bibinfo {author} {\bibfnamefont {Seongshik}\ \bibnamefont
  {Oh}}, \bibinfo {author} {\bibfnamefont {Tony~F}\ \bibnamefont {Heinz}},
  \bibinfo {author} {\bibfnamefont {David~A}\ \bibnamefont {Reis}},  \emph
  {et~al.},\ }\bibfield  {title} {\enquote {\bibinfo {title} {Probing
  topological phase transitions using high-harmonic generation},}\ }\href@noop
  {} {\bibfield  {journal} {\bibinfo  {journal} {Nature Photonics}\ }\textbf
  {\bibinfo {volume} {16}},\ \bibinfo {pages} {620--624} (\bibinfo {year}
  {2022})}\BibitemShut {NoStop}%
\bibitem [{\citenamefont {Bae}\ \emph {et~al.}(2022)\citenamefont {Bae},
  \citenamefont {Kim},\ and\ \citenamefont {Lee}}]{bae2022revealing}%
  \BibitemOpen
  \bibfield  {author} {\bibinfo {author} {\bibfnamefont {Gimin}\ \bibnamefont
  {Bae}}, \bibinfo {author} {\bibfnamefont {Youngjae}\ \bibnamefont {Kim}}, \
  and\ \bibinfo {author} {\bibfnamefont {JaeDong}\ \bibnamefont {Lee}},\
  }\bibfield  {title} {\enquote {\bibinfo {title} {Revealing berry curvature of
  the unoccupied band in high harmonic generation},}\ }\href@noop {} {\bibfield
   {journal} {\bibinfo  {journal} {Physical Review B}\ }\textbf {\bibinfo
  {volume} {106}},\ \bibinfo {pages} {205422} (\bibinfo {year}
  {2022})}\BibitemShut {NoStop}%
\bibitem [{\citenamefont {Rana}\ \emph {et~al.}(2022)\citenamefont {Rana},
  \citenamefont {Mrudul}, \citenamefont {Kartashov}, \citenamefont {Ivanov},\
  and\ \citenamefont {Dixit}}]{rana2022high}%
  \BibitemOpen
  \bibfield  {author} {\bibinfo {author} {\bibfnamefont {Navdeep}\ \bibnamefont
  {Rana}}, \bibinfo {author} {\bibfnamefont {MS}~\bibnamefont {Mrudul}},
  \bibinfo {author} {\bibfnamefont {Daniil}\ \bibnamefont {Kartashov}},
  \bibinfo {author} {\bibfnamefont {Misha}\ \bibnamefont {Ivanov}}, \ and\
  \bibinfo {author} {\bibfnamefont {Gopal}\ \bibnamefont {Dixit}},\ }\bibfield
  {title} {\enquote {\bibinfo {title} {High-harmonic spectroscopy of coherent
  lattice dynamics in graphene},}\ }\href@noop {} {\bibfield  {journal}
  {\bibinfo  {journal} {Physical Review B}\ }\textbf {\bibinfo {volume}
  {106}},\ \bibinfo {pages} {064303} (\bibinfo {year} {2022})}\BibitemShut
  {NoStop}%
\bibitem [{\citenamefont {Liu}\ \emph {et~al.}(2017)\citenamefont {Liu},
  \citenamefont {Li}, \citenamefont {You}, \citenamefont {Ghimire},
  \citenamefont {Heinz},\ and\ \citenamefont {Reis}}]{liu2017high}%
  \BibitemOpen
  \bibfield  {author} {\bibinfo {author} {\bibfnamefont {Hanzhe}\ \bibnamefont
  {Liu}}, \bibinfo {author} {\bibfnamefont {Yilei}\ \bibnamefont {Li}},
  \bibinfo {author} {\bibfnamefont {Yong~Sing}\ \bibnamefont {You}}, \bibinfo
  {author} {\bibfnamefont {Shambhu}\ \bibnamefont {Ghimire}}, \bibinfo {author}
  {\bibfnamefont {Tony~F}\ \bibnamefont {Heinz}}, \ and\ \bibinfo {author}
  {\bibfnamefont {David~A}\ \bibnamefont {Reis}},\ }\bibfield  {title}
  {\enquote {\bibinfo {title} {High-harmonic generation from an atomically thin
  semiconductor},}\ }\href@noop {} {\bibfield  {journal} {\bibinfo  {journal}
  {Nature Physics}\ }\textbf {\bibinfo {volume} {13}},\ \bibinfo {pages}
  {262--265} (\bibinfo {year} {2017})}\BibitemShut {NoStop}%
\bibitem [{\citenamefont {Silva}\ \emph {et~al.}(2018)\citenamefont {Silva},
  \citenamefont {Blinov}, \citenamefont {Rubtsov}, \citenamefont {Smirnova},\
  and\ \citenamefont {Ivanov}}]{silva2018high}%
  \BibitemOpen
  \bibfield  {author} {\bibinfo {author} {\bibfnamefont {REF}\ \bibnamefont
  {Silva}}, \bibinfo {author} {\bibfnamefont {Igor~V}\ \bibnamefont {Blinov}},
  \bibinfo {author} {\bibfnamefont {Alexey~N}\ \bibnamefont {Rubtsov}},
  \bibinfo {author} {\bibfnamefont {O}~\bibnamefont {Smirnova}}, \ and\
  \bibinfo {author} {\bibfnamefont {M}~\bibnamefont {Ivanov}},\ }\bibfield
  {title} {\enquote {\bibinfo {title} {High-harmonic spectroscopy of ultrafast
  many-body dynamics in strongly correlated systems},}\ }\href@noop {}
  {\bibfield  {journal} {\bibinfo  {journal} {Nature Photonics}\ }\textbf
  {\bibinfo {volume} {12}},\ \bibinfo {pages} {266--270} (\bibinfo {year}
  {2018})}\BibitemShut {NoStop}%
\bibitem [{\citenamefont {Murakami}\ \emph {et~al.}(2018)\citenamefont
  {Murakami}, \citenamefont {Eckstein},\ and\ \citenamefont
  {Werner}}]{murakami2018high}%
  \BibitemOpen
  \bibfield  {author} {\bibinfo {author} {\bibfnamefont {Yuta}\ \bibnamefont
  {Murakami}}, \bibinfo {author} {\bibfnamefont {Martin}\ \bibnamefont
  {Eckstein}}, \ and\ \bibinfo {author} {\bibfnamefont {Philipp}\ \bibnamefont
  {Werner}},\ }\bibfield  {title} {\enquote {\bibinfo {title} {High-harmonic
  generation in mott insulators},}\ }\href@noop {} {\bibfield  {journal}
  {\bibinfo  {journal} {Physical review letters}\ }\textbf {\bibinfo {volume}
  {121}},\ \bibinfo {pages} {057405} (\bibinfo {year} {2018})}\BibitemShut
  {NoStop}%
\bibitem [{\citenamefont {Takayoshi}\ \emph {et~al.}(2019)\citenamefont
  {Takayoshi}, \citenamefont {Murakami},\ and\ \citenamefont
  {Werner}}]{takayoshi2019high}%
  \BibitemOpen
  \bibfield  {author} {\bibinfo {author} {\bibfnamefont {Shintaro}\
  \bibnamefont {Takayoshi}}, \bibinfo {author} {\bibfnamefont {Yuta}\
  \bibnamefont {Murakami}}, \ and\ \bibinfo {author} {\bibfnamefont {Philipp}\
  \bibnamefont {Werner}},\ }\bibfield  {title} {\enquote {\bibinfo {title}
  {High-harmonic generation in quantum spin systems},}\ }\href@noop {}
  {\bibfield  {journal} {\bibinfo  {journal} {Physical Review B}\ }\textbf
  {\bibinfo {volume} {99}},\ \bibinfo {pages} {184303} (\bibinfo {year}
  {2019})}\BibitemShut {NoStop}%
\bibitem [{\citenamefont {Imai}\ \emph {et~al.}(2020)\citenamefont {Imai},
  \citenamefont {Ono},\ and\ \citenamefont {Ishihara}}]{imai2020high}%
  \BibitemOpen
  \bibfield  {author} {\bibinfo {author} {\bibfnamefont {Shohei}\ \bibnamefont
  {Imai}}, \bibinfo {author} {\bibfnamefont {Atsushi}\ \bibnamefont {Ono}}, \
  and\ \bibinfo {author} {\bibfnamefont {Sumio}\ \bibnamefont {Ishihara}},\
  }\bibfield  {title} {\enquote {\bibinfo {title} {High harmonic generation in
  a correlated electron system},}\ }\href@noop {} {\bibfield  {journal}
  {\bibinfo  {journal} {Physical review letters}\ }\textbf {\bibinfo {volume}
  {124}},\ \bibinfo {pages} {157404} (\bibinfo {year} {2020})}\BibitemShut
  {NoStop}%
\bibitem [{\citenamefont {Lysne}\ \emph {et~al.}(2020)\citenamefont {Lysne},
  \citenamefont {Murakami},\ and\ \citenamefont
  {Werner}}]{lysne2020signatures}%
  \BibitemOpen
  \bibfield  {author} {\bibinfo {author} {\bibfnamefont {Markus}\ \bibnamefont
  {Lysne}}, \bibinfo {author} {\bibfnamefont {Yuta}\ \bibnamefont {Murakami}},
  \ and\ \bibinfo {author} {\bibfnamefont {Philipp}\ \bibnamefont {Werner}},\
  }\bibfield  {title} {\enquote {\bibinfo {title} {Signatures of bosonic
  excitations in high-harmonic spectra of mott insulators},}\ }\href@noop {}
  {\bibfield  {journal} {\bibinfo  {journal} {Physical Review B}\ }\textbf
  {\bibinfo {volume} {101}},\ \bibinfo {pages} {195139} (\bibinfo {year}
  {2020})}\BibitemShut {NoStop}%
\bibitem [{\citenamefont {Murakami}\ \emph {et~al.}(2021)\citenamefont
  {Murakami}, \citenamefont {Takayoshi}, \citenamefont {Koga},\ and\
  \citenamefont {Werner}}]{murakami2021high}%
  \BibitemOpen
  \bibfield  {author} {\bibinfo {author} {\bibfnamefont {Yuta}\ \bibnamefont
  {Murakami}}, \bibinfo {author} {\bibfnamefont {Shintaro}\ \bibnamefont
  {Takayoshi}}, \bibinfo {author} {\bibfnamefont {Akihisa}\ \bibnamefont
  {Koga}}, \ and\ \bibinfo {author} {\bibfnamefont {Philipp}\ \bibnamefont
  {Werner}},\ }\bibfield  {title} {\enquote {\bibinfo {title} {High-harmonic
  generation in one-dimensional mott insulators},}\ }\href@noop {} {\bibfield
  {journal} {\bibinfo  {journal} {Physical Review B}\ }\textbf {\bibinfo
  {volume} {103}},\ \bibinfo {pages} {035110} (\bibinfo {year}
  {2021})}\BibitemShut {NoStop}%
\bibitem [{\citenamefont {Zhu}\ \emph {et~al.}(2021)\citenamefont {Zhu},
  \citenamefont {Fauseweh}, \citenamefont {Chacon},\ and\ \citenamefont
  {Zhu}}]{zhu2021ultrafast}%
  \BibitemOpen
  \bibfield  {author} {\bibinfo {author} {\bibfnamefont {Wei}\ \bibnamefont
  {Zhu}}, \bibinfo {author} {\bibfnamefont {Benedikt}\ \bibnamefont
  {Fauseweh}}, \bibinfo {author} {\bibfnamefont {Alexis}\ \bibnamefont
  {Chacon}}, \ and\ \bibinfo {author} {\bibfnamefont {Jian-Xin}\ \bibnamefont
  {Zhu}},\ }\bibfield  {title} {\enquote {\bibinfo {title} {Ultrafast
  laser-driven many-body dynamics and kondo coherence collapse},}\ }\href@noop
  {} {\bibfield  {journal} {\bibinfo  {journal} {Physical Review B}\ }\textbf
  {\bibinfo {volume} {103}},\ \bibinfo {pages} {224305} (\bibinfo {year}
  {2021})}\BibitemShut {NoStop}%
\bibitem [{\citenamefont {Orthodoxou}\ \emph {et~al.}(2021)\citenamefont
  {Orthodoxou}, \citenamefont {Za{\"\i}r},\ and\ \citenamefont
  {Booth}}]{orthodoxou2021high}%
  \BibitemOpen
  \bibfield  {author} {\bibinfo {author} {\bibfnamefont {Christopher}\
  \bibnamefont {Orthodoxou}}, \bibinfo {author} {\bibfnamefont {Amelle}\
  \bibnamefont {Za{\"\i}r}}, \ and\ \bibinfo {author} {\bibfnamefont
  {George~H}\ \bibnamefont {Booth}},\ }\bibfield  {title} {\enquote {\bibinfo
  {title} {High harmonic generation in two-dimensional mott insulators},}\
  }\href@noop {} {\bibfield  {journal} {\bibinfo  {journal} {npj Quantum
  Materials}\ }\textbf {\bibinfo {volume} {6}},\ \bibinfo {pages} {76}
  (\bibinfo {year} {2021})}\BibitemShut {NoStop}%
\bibitem [{\citenamefont {Shao}\ \emph {et~al.}(2022)\citenamefont {Shao},
  \citenamefont {Lu}, \citenamefont {Zhang}, \citenamefont {Yu}, \citenamefont
  {Tohyama},\ and\ \citenamefont {Lu}}]{shao2022high}%
  \BibitemOpen
  \bibfield  {author} {\bibinfo {author} {\bibfnamefont {Can}\ \bibnamefont
  {Shao}}, \bibinfo {author} {\bibfnamefont {Hantao}\ \bibnamefont {Lu}},
  \bibinfo {author} {\bibfnamefont {Xiao}\ \bibnamefont {Zhang}}, \bibinfo
  {author} {\bibfnamefont {Chao}\ \bibnamefont {Yu}}, \bibinfo {author}
  {\bibfnamefont {Takami}\ \bibnamefont {Tohyama}}, \ and\ \bibinfo {author}
  {\bibfnamefont {Ruifeng}\ \bibnamefont {Lu}},\ }\bibfield  {title} {\enquote
  {\bibinfo {title} {High-harmonic generation approaching the quantum critical
  point of strongly correlated systems},}\ }\href@noop {} {\bibfield  {journal}
  {\bibinfo  {journal} {Physical review letters}\ }\textbf {\bibinfo {volume}
  {128}},\ \bibinfo {pages} {047401} (\bibinfo {year} {2022})}\BibitemShut
  {NoStop}%
\bibitem [{\citenamefont {Alcal{\`a}}\ \emph {et~al.}(2022)\citenamefont
  {Alcal{\`a}}, \citenamefont {Bhattacharya}, \citenamefont {Biegert},
  \citenamefont {Ciappina}, \citenamefont {Elu}, \citenamefont {Gra{\ss}},
  \citenamefont {Grochowski}, \citenamefont {Lewenstein}, \citenamefont
  {Palau}, \citenamefont {Sidiropoulos} \emph {et~al.}}]{alcala2022high}%
  \BibitemOpen
  \bibfield  {author} {\bibinfo {author} {\bibfnamefont {Jordi}\ \bibnamefont
  {Alcal{\`a}}}, \bibinfo {author} {\bibfnamefont {Utso}\ \bibnamefont
  {Bhattacharya}}, \bibinfo {author} {\bibfnamefont {Jens}\ \bibnamefont
  {Biegert}}, \bibinfo {author} {\bibfnamefont {Marcelo}\ \bibnamefont
  {Ciappina}}, \bibinfo {author} {\bibfnamefont {Ugaitz}\ \bibnamefont {Elu}},
  \bibinfo {author} {\bibfnamefont {Tobias}\ \bibnamefont {Gra{\ss}}}, \bibinfo
  {author} {\bibfnamefont {Piotr~T}\ \bibnamefont {Grochowski}}, \bibinfo
  {author} {\bibfnamefont {Maciej}\ \bibnamefont {Lewenstein}}, \bibinfo
  {author} {\bibfnamefont {Anna}\ \bibnamefont {Palau}}, \bibinfo {author}
  {\bibfnamefont {Themistoklis~PH}\ \bibnamefont {Sidiropoulos}},  \emph
  {et~al.},\ }\bibfield  {title} {\enquote {\bibinfo {title} {High-harmonic
  spectroscopy of quantum phase transitions in a high-tc superconductor},}\
  }\href@noop {} {\bibfield  {journal} {\bibinfo  {journal} {Proceedings of the
  National Academy of Sciences}\ }\textbf {\bibinfo {volume} {119}},\ \bibinfo
  {pages} {e2207766119} (\bibinfo {year} {2022})}\BibitemShut {NoStop}%
\bibitem [{\citenamefont {Hansen}\ \emph {et~al.}(2022)\citenamefont {Hansen},
  \citenamefont {Jensen},\ and\ \citenamefont
  {Madsen}}]{hansen2022correlation}%
  \BibitemOpen
  \bibfield  {author} {\bibinfo {author} {\bibfnamefont {Thomas}\ \bibnamefont
  {Hansen}}, \bibinfo {author} {\bibfnamefont {Simon Vendelbo~Bylling}\
  \bibnamefont {Jensen}}, \ and\ \bibinfo {author} {\bibfnamefont {Lars~Bojer}\
  \bibnamefont {Madsen}},\ }\bibfield  {title} {\enquote {\bibinfo {title}
  {Correlation effects in high-order harmonic generation from finite
  systems},}\ }\href@noop {} {\bibfield  {journal} {\bibinfo  {journal}
  {Physical Review A}\ }\textbf {\bibinfo {volume} {105}},\ \bibinfo {pages}
  {053118} (\bibinfo {year} {2022})}\BibitemShut {NoStop}%
\bibitem [{\citenamefont {Uchida}\ \emph {et~al.}(2022)\citenamefont {Uchida},
  \citenamefont {Mattoni}, \citenamefont {Yonezawa}, \citenamefont {Nakamura},
  \citenamefont {Maeno},\ and\ \citenamefont {Tanaka}}]{uchida2022high}%
  \BibitemOpen
  \bibfield  {author} {\bibinfo {author} {\bibfnamefont {K}~\bibnamefont
  {Uchida}}, \bibinfo {author} {\bibfnamefont {G}~\bibnamefont {Mattoni}},
  \bibinfo {author} {\bibfnamefont {S}~\bibnamefont {Yonezawa}}, \bibinfo
  {author} {\bibfnamefont {F}~\bibnamefont {Nakamura}}, \bibinfo {author}
  {\bibfnamefont {Y}~\bibnamefont {Maeno}}, \ and\ \bibinfo {author}
  {\bibfnamefont {K}~\bibnamefont {Tanaka}},\ }\bibfield  {title} {\enquote
  {\bibinfo {title} {High-order harmonic generation and its unconventional
  scaling law in the mott-insulating ca 2 ruo 4},}\ }\href@noop {} {\bibfield
  {journal} {\bibinfo  {journal} {Physical Review Letters}\ }\textbf {\bibinfo
  {volume} {128}},\ \bibinfo {pages} {127401} (\bibinfo {year}
  {2022})}\BibitemShut {NoStop}%
\bibitem [{\citenamefont {AlShafey}\ \emph {et~al.}(2022)\citenamefont
  {AlShafey}, \citenamefont {McCaul}, \citenamefont {Lu}, \citenamefont {Jia},
  \citenamefont {Gong}, \citenamefont {Addison}, \citenamefont {Bondar},
  \citenamefont {Randeria},\ and\ \citenamefont
  {Landsman}}]{alshafey2022ultrafast}%
  \BibitemOpen
  \bibfield  {author} {\bibinfo {author} {\bibfnamefont {Abdallah}\
  \bibnamefont {AlShafey}}, \bibinfo {author} {\bibfnamefont {Gerard}\
  \bibnamefont {McCaul}}, \bibinfo {author} {\bibfnamefont {Yuan-Ming}\
  \bibnamefont {Lu}}, \bibinfo {author} {\bibfnamefont {Xu-Yan}\ \bibnamefont
  {Jia}}, \bibinfo {author} {\bibfnamefont {Shou-Shu}\ \bibnamefont {Gong}},
  \bibinfo {author} {\bibfnamefont {Zachariah}\ \bibnamefont {Addison}},
  \bibinfo {author} {\bibfnamefont {Denys~I}\ \bibnamefont {Bondar}}, \bibinfo
  {author} {\bibfnamefont {Mohit}\ \bibnamefont {Randeria}}, \ and\ \bibinfo
  {author} {\bibfnamefont {Alexandra~S}\ \bibnamefont {Landsman}},\ }\bibfield
  {title} {\enquote {\bibinfo {title} {Ultrafast laser-driven dynamics in
  metal-insulator interface},}\ }\href@noop {} {\bibfield  {journal} {\bibinfo
  {journal} {arXiv preprint arXiv:2212.09176}\ } (\bibinfo {year}
  {2022})}\BibitemShut {NoStop}%
\bibitem [{\citenamefont {Pizzi}\ \emph {et~al.}(2023)\citenamefont {Pizzi},
  \citenamefont {Gorlach}, \citenamefont {Rivera}, \citenamefont {Nunnenkamp},\
  and\ \citenamefont {Kaminer}}]{pizzi2023light}%
  \BibitemOpen
  \bibfield  {author} {\bibinfo {author} {\bibfnamefont {Andrea}\ \bibnamefont
  {Pizzi}}, \bibinfo {author} {\bibfnamefont {Alexey}\ \bibnamefont {Gorlach}},
  \bibinfo {author} {\bibfnamefont {Nicholas}\ \bibnamefont {Rivera}}, \bibinfo
  {author} {\bibfnamefont {Andreas}\ \bibnamefont {Nunnenkamp}}, \ and\
  \bibinfo {author} {\bibfnamefont {Ido}\ \bibnamefont {Kaminer}},\ }\bibfield
  {title} {\enquote {\bibinfo {title} {Light emission from strongly driven
  many-body systems},}\ }\href@noop {} {\bibfield  {journal} {\bibinfo
  {journal} {Nature Physics}\ ,\ \bibinfo {pages} {1--11}} (\bibinfo {year}
  {2023})}\BibitemShut {NoStop}%
\bibitem [{\citenamefont {Shimomura}\ \emph {et~al.}(2023)\citenamefont
  {Shimomura}, \citenamefont {Uchida}, \citenamefont {Nagai}, \citenamefont
  {Kusaba},\ and\ \citenamefont {Tanaka}}]{shimomura2023ultrafast}%
  \BibitemOpen
  \bibfield  {author} {\bibinfo {author} {\bibfnamefont {KS}~\bibnamefont
  {Shimomura}}, \bibinfo {author} {\bibfnamefont {K}~\bibnamefont {Uchida}},
  \bibinfo {author} {\bibfnamefont {N}~\bibnamefont {Nagai}}, \bibinfo {author}
  {\bibfnamefont {S}~\bibnamefont {Kusaba}}, \ and\ \bibinfo {author}
  {\bibfnamefont {K}~\bibnamefont {Tanaka}},\ }\bibfield  {title} {\enquote
  {\bibinfo {title} {Ultrafast electron-electron scattering in metallic phase
  of 2h-nbse $ \_2 $ probed by high harmonic generation},}\ }\href@noop {}
  {\bibfield  {journal} {\bibinfo  {journal} {arXiv preprint arXiv:2302.04984}\
  } (\bibinfo {year} {2023})}\BibitemShut {NoStop}%
\bibitem [{\citenamefont {Corkum}(1993)}]{Corkum1993}%
  \BibitemOpen
  \bibfield  {author} {\bibinfo {author} {\bibfnamefont {P.~B.}\ \bibnamefont
  {Corkum}},\ }\bibfield  {title} {\enquote {\bibinfo {title} {Plasma
  perspective on strong field multiphoton ionization},}\ }\href {\doibase
  10.1103/PhysRevLett.71.1994} {\bibfield  {journal} {\bibinfo  {journal}
  {Phys. Rev. Lett.}\ }\textbf {\bibinfo {volume} {71}},\ \bibinfo {pages}
  {1994--1997} (\bibinfo {year} {1993})}\BibitemShut {NoStop}%
\bibitem [{\citenamefont {Lewenstein}\ \emph {et~al.}(1994)\citenamefont
  {Lewenstein}, \citenamefont {Balcou}, \citenamefont {Ivanov}, \citenamefont
  {L’huillier},\ and\ \citenamefont {Corkum}}]{lewenstein1994theory}%
  \BibitemOpen
  \bibfield  {author} {\bibinfo {author} {\bibfnamefont {Maciej}\ \bibnamefont
  {Lewenstein}}, \bibinfo {author} {\bibfnamefont {Ph}~\bibnamefont {Balcou}},
  \bibinfo {author} {\bibfnamefont {M~Yu}\ \bibnamefont {Ivanov}}, \bibinfo
  {author} {\bibfnamefont {Anne}\ \bibnamefont {L’huillier}}, \ and\ \bibinfo
  {author} {\bibfnamefont {Paul~B}\ \bibnamefont {Corkum}},\ }\bibfield
  {title} {\enquote {\bibinfo {title} {Theory of high-harmonic generation by
  low-frequency laser fields},}\ }\href@noop {} {\bibfield  {journal} {\bibinfo
   {journal} {Physical Review A}\ }\textbf {\bibinfo {volume} {49}},\ \bibinfo
  {pages} {2117} (\bibinfo {year} {1994})}\BibitemShut {NoStop}%
\bibitem [{\citenamefont {Vampa}\ \emph
  {et~al.}(2015{\natexlab{b}})\citenamefont {Vampa}, \citenamefont {McDonald},
  \citenamefont {Orlando}, \citenamefont {Corkum},\ and\ \citenamefont
  {Brabec}}]{vampa2015}%
  \BibitemOpen
  \bibfield  {author} {\bibinfo {author} {\bibfnamefont {G.}~\bibnamefont
  {Vampa}}, \bibinfo {author} {\bibfnamefont {C.~R.}\ \bibnamefont {McDonald}},
  \bibinfo {author} {\bibfnamefont {G.}~\bibnamefont {Orlando}}, \bibinfo
  {author} {\bibfnamefont {P.~B.}\ \bibnamefont {Corkum}}, \ and\ \bibinfo
  {author} {\bibfnamefont {T.}~\bibnamefont {Brabec}},\ }\bibfield  {title}
  {\enquote {\bibinfo {title} {Semiclassical analysis of high harmonic
  generation in bulk crystals},}\ }\href {\doibase 10.1103/PhysRevB.91.064302}
  {\bibfield  {journal} {\bibinfo  {journal} {Phys. Rev. B}\ }\textbf {\bibinfo
  {volume} {91}},\ \bibinfo {pages} {064302} (\bibinfo {year}
  {2015}{\natexlab{b}})}\BibitemShut {NoStop}%
\bibitem [{\citenamefont {Yoshikawa}\ \emph {et~al.}(2017)\citenamefont
  {Yoshikawa}, \citenamefont {Tamaya},\ and\ \citenamefont
  {Tanaka}}]{yoshikawa2017high}%
  \BibitemOpen
  \bibfield  {author} {\bibinfo {author} {\bibfnamefont {Naotaka}\ \bibnamefont
  {Yoshikawa}}, \bibinfo {author} {\bibfnamefont {Tomohiro}\ \bibnamefont
  {Tamaya}}, \ and\ \bibinfo {author} {\bibfnamefont {Koichiro}\ \bibnamefont
  {Tanaka}},\ }\bibfield  {title} {\enquote {\bibinfo {title} {High-harmonic
  generation in graphene enhanced by elliptically polarized light
  excitation},}\ }\href@noop {} {\bibfield  {journal} {\bibinfo  {journal}
  {Science}\ }\textbf {\bibinfo {volume} {356}},\ \bibinfo {pages} {736--738}
  (\bibinfo {year} {2017})}\BibitemShut {NoStop}%
\bibitem [{\citenamefont {Ishii}\ \emph {et~al.}(2014)\citenamefont {Ishii},
  \citenamefont {Kaneshima}, \citenamefont {Kitano}, \citenamefont {Kanai},
  \citenamefont {Watanabe},\ and\ \citenamefont {Itatani}}]{Ishii2014}%
  \BibitemOpen
  \bibfield  {author} {\bibinfo {author} {\bibfnamefont {Nobuhisa}\
  \bibnamefont {Ishii}}, \bibinfo {author} {\bibfnamefont {Keisuke}\
  \bibnamefont {Kaneshima}}, \bibinfo {author} {\bibfnamefont {Kenta}\
  \bibnamefont {Kitano}}, \bibinfo {author} {\bibfnamefont {Teruto}\
  \bibnamefont {Kanai}}, \bibinfo {author} {\bibfnamefont {Shuntaro}\
  \bibnamefont {Watanabe}}, \ and\ \bibinfo {author} {\bibfnamefont {Jiro}\
  \bibnamefont {Itatani}},\ }\bibfield  {title} {\enquote {\bibinfo {title}
  {Carrier-envelope phase-dependent high harmonic generation in the water
  window using few-cycle infrared pulses},}\ }\href {\doibase
  10.1038/ncomms4331} {\bibfield  {journal} {\bibinfo  {journal} {Nature
  Communications}\ }\textbf {\bibinfo {volume} {5}},\ \bibinfo {pages} {3331}
  (\bibinfo {year} {2014})}\BibitemShut {NoStop}%
\bibitem [{\citenamefont {Murakami}\ \emph {et~al.}(2022)\citenamefont
  {Murakami}, \citenamefont {Uchida}, \citenamefont {Koga}, \citenamefont
  {Tanaka},\ and\ \citenamefont {Werner}}]{Murakami2022anomalous}%
  \BibitemOpen
  \bibfield  {author} {\bibinfo {author} {\bibfnamefont {Yuta}\ \bibnamefont
  {Murakami}}, \bibinfo {author} {\bibfnamefont {Kento}\ \bibnamefont
  {Uchida}}, \bibinfo {author} {\bibfnamefont {Akihisa}\ \bibnamefont {Koga}},
  \bibinfo {author} {\bibfnamefont {Koichiro}\ \bibnamefont {Tanaka}}, \ and\
  \bibinfo {author} {\bibfnamefont {Philipp}\ \bibnamefont {Werner}},\
  }\bibfield  {title} {\enquote {\bibinfo {title} {Anomalous temperature
  dependence of high-harmonic generation in mott insulators},}\ }\href
  {\doibase 10.1103/PhysRevLett.129.157401} {\bibfield  {journal} {\bibinfo
  {journal} {Phys. Rev. Lett.}\ }\textbf {\bibinfo {volume} {129}},\ \bibinfo
  {pages} {157401} (\bibinfo {year} {2022})}\BibitemShut {NoStop}%
\bibitem [{\citenamefont {Murakami}\ and\ \citenamefont
  {Sch{\"u}ler}(2022)}]{murakami2022doping}%
  \BibitemOpen
  \bibfield  {author} {\bibinfo {author} {\bibfnamefont {Yuta}\ \bibnamefont
  {Murakami}}\ and\ \bibinfo {author} {\bibfnamefont {Michael}\ \bibnamefont
  {Sch{\"u}ler}},\ }\bibfield  {title} {\enquote {\bibinfo {title} {Doping and
  gap size dependence of high-harmonic generation in graphene: Importance of
  consistent formulation of light-matter coupling},}\ }\href@noop {} {\bibfield
   {journal} {\bibinfo  {journal} {Physical Review B}\ }\textbf {\bibinfo
  {volume} {106}},\ \bibinfo {pages} {035204} (\bibinfo {year}
  {2022})}\BibitemShut {NoStop}%
\bibitem [{\citenamefont {Kaplan}\ and\ \citenamefont
  {Shkolnikov}(1994)}]{kaplan1994superdressed}%
  \BibitemOpen
  \bibfield  {author} {\bibinfo {author} {\bibfnamefont {AE}~\bibnamefont
  {Kaplan}}\ and\ \bibinfo {author} {\bibfnamefont {PL}~\bibnamefont
  {Shkolnikov}},\ }\bibfield  {title} {\enquote {\bibinfo {title} {Superdressed
  two-level atom: very high harmonic generation and multiresonances},}\
  }\href@noop {} {\bibfield  {journal} {\bibinfo  {journal} {Physical Review
  A}\ }\textbf {\bibinfo {volume} {49}},\ \bibinfo {pages} {1275} (\bibinfo
  {year} {1994})}\BibitemShut {NoStop}%
\bibitem [{\citenamefont {Krainov}\ and\ \citenamefont
  {Mulyukov}(1994)}]{krainov1994plateau}%
  \BibitemOpen
  \bibfield  {author} {\bibinfo {author} {\bibfnamefont {VP}~\bibnamefont
  {Krainov}}\ and\ \bibinfo {author} {\bibfnamefont {ZS}~\bibnamefont
  {Mulyukov}},\ }\bibfield  {title} {\enquote {\bibinfo {title} {A plateau in
  high-order harmonic generation for a two-level atom},}\ }\href@noop {}
  {\bibfield  {journal} {\bibinfo  {journal} {Laser Phys}\ }\textbf {\bibinfo
  {volume} {4}},\ \bibinfo {pages} {544} (\bibinfo {year} {1994})}\BibitemShut
  {NoStop}%
\bibitem [{\citenamefont {de~Morisson~Faria}\ and\ \citenamefont
  {Rotter}(2002)}]{de2002high}%
  \BibitemOpen
  \bibfield  {author} {\bibinfo {author} {\bibfnamefont {C~Figueira}\
  \bibnamefont {de~Morisson~Faria}}\ and\ \bibinfo {author} {\bibfnamefont
  {I}~\bibnamefont {Rotter}},\ }\bibfield  {title} {\enquote {\bibinfo {title}
  {High-order harmonic generation in a driven two-level atom: Periodic level
  crossings and three-step processes},}\ }\href@noop {} {\bibfield  {journal}
  {\bibinfo  {journal} {Physical Review A}\ }\textbf {\bibinfo {volume} {66}},\
  \bibinfo {pages} {013402} (\bibinfo {year} {2002})}\BibitemShut {NoStop}%
\bibitem [{\citenamefont {Boyd}(2020)}]{boyd2020nonlinear}%
  \BibitemOpen
  \bibfield  {author} {\bibinfo {author} {\bibfnamefont {Robert~W}\
  \bibnamefont {Boyd}},\ }\href@noop {} {\emph {\bibinfo {title} {Nonlinear
  optics}}}\ (\bibinfo  {publisher} {Academic press},\ \bibinfo {year}
  {2020})\BibitemShut {NoStop}%
\bibitem [{\citenamefont {Birnir}\ \emph {et~al.}(1993)\citenamefont {Birnir},
  \citenamefont {Galdrikian}, \citenamefont {Grauer},\ and\ \citenamefont
  {Sherwin}}]{birnir1993nonperturbative}%
  \BibitemOpen
  \bibfield  {author} {\bibinfo {author} {\bibfnamefont {Bj{\"o}rn}\
  \bibnamefont {Birnir}}, \bibinfo {author} {\bibfnamefont {Bryan}\
  \bibnamefont {Galdrikian}}, \bibinfo {author} {\bibfnamefont {Rainer}\
  \bibnamefont {Grauer}}, \ and\ \bibinfo {author} {\bibfnamefont {Mark}\
  \bibnamefont {Sherwin}},\ }\bibfield  {title} {\enquote {\bibinfo {title}
  {Nonperturbative resonances in periodically driven quantum wells},}\
  }\href@noop {} {\bibfield  {journal} {\bibinfo  {journal} {Physical Review
  B}\ }\textbf {\bibinfo {volume} {47}},\ \bibinfo {pages} {6795} (\bibinfo
  {year} {1993})}\BibitemShut {NoStop}%
\bibitem [{\citenamefont {Nikonov}\ \emph {et~al.}(1997)\citenamefont
  {Nikonov}, \citenamefont {Imamo{\u{g}}lu}, \citenamefont {Butov},\ and\
  \citenamefont {Schmidt}}]{nikonov1997collective}%
  \BibitemOpen
  \bibfield  {author} {\bibinfo {author} {\bibfnamefont {Dmitri~E}\
  \bibnamefont {Nikonov}}, \bibinfo {author} {\bibfnamefont {Ata{\c{c}}}\
  \bibnamefont {Imamo{\u{g}}lu}}, \bibinfo {author} {\bibfnamefont {Leonid~V}\
  \bibnamefont {Butov}}, \ and\ \bibinfo {author} {\bibfnamefont {Holger}\
  \bibnamefont {Schmidt}},\ }\bibfield  {title} {\enquote {\bibinfo {title}
  {Collective intersubband excitations in quantum wells: Coulomb interaction
  versus subband dispersion},}\ }\href@noop {} {\bibfield  {journal} {\bibinfo
  {journal} {Physical review letters}\ }\textbf {\bibinfo {volume} {79}},\
  \bibinfo {pages} {4633} (\bibinfo {year} {1997})}\BibitemShut {NoStop}%
\bibitem [{\citenamefont {Kishida}\ \emph {et~al.}(2000)\citenamefont
  {Kishida}, \citenamefont {Matsuzaki}, \citenamefont {Okamoto}, \citenamefont
  {Manabe}, \citenamefont {Yamashita}, \citenamefont {Taguchi},\ and\
  \citenamefont {Tokura}}]{kishida2000gigantic}%
  \BibitemOpen
  \bibfield  {author} {\bibinfo {author} {\bibfnamefont {H}~\bibnamefont
  {Kishida}}, \bibinfo {author} {\bibfnamefont {H}~\bibnamefont {Matsuzaki}},
  \bibinfo {author} {\bibfnamefont {H}~\bibnamefont {Okamoto}}, \bibinfo
  {author} {\bibfnamefont {T}~\bibnamefont {Manabe}}, \bibinfo {author}
  {\bibfnamefont {M}~\bibnamefont {Yamashita}}, \bibinfo {author}
  {\bibfnamefont {Y}~\bibnamefont {Taguchi}}, \ and\ \bibinfo {author}
  {\bibfnamefont {Y}~\bibnamefont {Tokura}},\ }\bibfield  {title} {\enquote
  {\bibinfo {title} {Gigantic optical nonlinearity in one-dimensional
  mott--hubbard insulators},}\ }\href@noop {} {\bibfield  {journal} {\bibinfo
  {journal} {Nature}\ }\textbf {\bibinfo {volume} {405}},\ \bibinfo {pages}
  {929--932} (\bibinfo {year} {2000})}\BibitemShut {NoStop}%
\bibitem [{\citenamefont {Ogasawara}\ \emph {et~al.}(2000)\citenamefont
  {Ogasawara}, \citenamefont {Ashida}, \citenamefont {Motoyama}, \citenamefont
  {Eisaki}, \citenamefont {Uchida}, \citenamefont {Tokura}, \citenamefont
  {Ghosh}, \citenamefont {Shukla}, \citenamefont {Mazumdar},\ and\
  \citenamefont {Kuwata-Gonokami}}]{ogasawara2000ultrafast}%
  \BibitemOpen
  \bibfield  {author} {\bibinfo {author} {\bibfnamefont {T}~\bibnamefont
  {Ogasawara}}, \bibinfo {author} {\bibfnamefont {M}~\bibnamefont {Ashida}},
  \bibinfo {author} {\bibfnamefont {N}~\bibnamefont {Motoyama}}, \bibinfo
  {author} {\bibfnamefont {H}~\bibnamefont {Eisaki}}, \bibinfo {author}
  {\bibfnamefont {S}~\bibnamefont {Uchida}}, \bibinfo {author} {\bibfnamefont
  {Y}~\bibnamefont {Tokura}}, \bibinfo {author} {\bibfnamefont {H}~\bibnamefont
  {Ghosh}}, \bibinfo {author} {\bibfnamefont {A}~\bibnamefont {Shukla}},
  \bibinfo {author} {\bibfnamefont {S}~\bibnamefont {Mazumdar}}, \ and\
  \bibinfo {author} {\bibfnamefont {M}~\bibnamefont {Kuwata-Gonokami}},\
  }\bibfield  {title} {\enquote {\bibinfo {title} {Ultrafast optical
  nonlinearity in the quasi-one-dimensional mott insulator sr 2 cuo 3},}\
  }\href@noop {} {\bibfield  {journal} {\bibinfo  {journal} {Physical review
  letters}\ }\textbf {\bibinfo {volume} {85}},\ \bibinfo {pages} {2204}
  (\bibinfo {year} {2000})}\BibitemShut {NoStop}%
\bibitem [{\citenamefont {Mizuno}\ \emph {et~al.}(2000)\citenamefont {Mizuno},
  \citenamefont {Tsutsui}, \citenamefont {Tohyama},\ and\ \citenamefont
  {Maekawa}}]{mizuno2000nonlinear}%
  \BibitemOpen
  \bibfield  {author} {\bibinfo {author} {\bibfnamefont {Y}~\bibnamefont
  {Mizuno}}, \bibinfo {author} {\bibfnamefont {K}~\bibnamefont {Tsutsui}},
  \bibinfo {author} {\bibfnamefont {T}~\bibnamefont {Tohyama}}, \ and\ \bibinfo
  {author} {\bibfnamefont {S}~\bibnamefont {Maekawa}},\ }\bibfield  {title}
  {\enquote {\bibinfo {title} {Nonlinear optical response and spin-charge
  separation in one-dimensional mott insulators},}\ }\href@noop {} {\bibfield
  {journal} {\bibinfo  {journal} {Physical Review B}\ }\textbf {\bibinfo
  {volume} {62}},\ \bibinfo {pages} {R4769} (\bibinfo {year}
  {2000})}\BibitemShut {NoStop}%
\bibitem [{\citenamefont {Zhang}(2001)}]{zhang2001origin}%
  \BibitemOpen
  \bibfield  {author} {\bibinfo {author} {\bibfnamefont {GP}~\bibnamefont
  {Zhang}},\ }\bibfield  {title} {\enquote {\bibinfo {title} {Origin of giant
  optical nonlinearity in charge-transfer--mott insulators: A new paradigm for
  nonlinear optics},}\ }\href@noop {} {\bibfield  {journal} {\bibinfo
  {journal} {Physical Review Letters}\ }\textbf {\bibinfo {volume} {86}},\
  \bibinfo {pages} {2086} (\bibinfo {year} {2001})}\BibitemShut {NoStop}%
\bibitem [{\citenamefont {Sarantseva}\ \emph {et~al.}(2021)\citenamefont
  {Sarantseva}, \citenamefont {Silaev}, \citenamefont {Romanov}, \citenamefont
  {Vvedenskii},\ and\ \citenamefont {Frolov}}]{sarantseva2021time}%
  \BibitemOpen
  \bibfield  {author} {\bibinfo {author} {\bibfnamefont {TS}~\bibnamefont
  {Sarantseva}}, \bibinfo {author} {\bibfnamefont {AA}~\bibnamefont {Silaev}},
  \bibinfo {author} {\bibfnamefont {AA}~\bibnamefont {Romanov}}, \bibinfo
  {author} {\bibfnamefont {NV}~\bibnamefont {Vvedenskii}}, \ and\ \bibinfo
  {author} {\bibfnamefont {MV}~\bibnamefont {Frolov}},\ }\bibfield  {title}
  {\enquote {\bibinfo {title} {Time-frequency analysis of high harmonic
  generation using a probe xuv pulse},}\ }\href@noop {} {\bibfield  {journal}
  {\bibinfo  {journal} {Optics Express}\ }\textbf {\bibinfo {volume} {29}},\
  \bibinfo {pages} {1428--1440} (\bibinfo {year} {2021})}\BibitemShut {NoStop}%
\bibitem [{\citenamefont {Haug}\ and\ \citenamefont
  {Koch}(2009)}]{haug2009quantum}%
  \BibitemOpen
  \bibfield  {author} {\bibinfo {author} {\bibfnamefont {Hartmut}\ \bibnamefont
  {Haug}}\ and\ \bibinfo {author} {\bibfnamefont {Stephan~W}\ \bibnamefont
  {Koch}},\ }\href@noop {} {\emph {\bibinfo {title} {Quantum theory of the
  optical and electronic properties of semiconductors}}}\ (\bibinfo
  {publisher} {World Scientific Publishing Company},\ \bibinfo {year}
  {2009})\BibitemShut {NoStop}%
\end{thebibliography}%

\end{document}